\documentclass
[aps,reprint,showpacs,floatfix,amsmath,amssymb,superscriptaddress,longbibliography,footinbib]
{revtex4-2}

\usepackage{graphicx}
\usepackage{dcolumn}
\usepackage{bm}
\usepackage[colorlinks=true,urlcolor=blue,citecolor=blue,linkcolor=blue]{hyperref}
\usepackage{braket}
\usepackage{comment}
\usepackage{mathtools}
\usepackage{multirow}
\usepackage{lineno}
\usepackage{siunitx}
\usepackage{orcidlink}
\graphicspath{{./Fig/}}

\newcommand{\g}{^1S_0}
\newcommand{\e}{4f^{13}5d6s^2 \: (J=2)}
\newcommand{\clock}{^1S_0 \leftrightarrow 4f^{13}5d6s^2 \: (J=2)}
\newcommand{\gtoe}{\ket{g} \leftrightarrow \ket{e}}
\newcommand{\dn}[2]{\delta \nu_{#1}^{#2}}

\newcommand{\w}[1]{w^{#1}}
\newcommand{\dr}[1]{\delta \langle r^2 \rangle ^{#1}}
\newcommand{\drr}[1]{\delta \langle r^4 \rangle ^{#1}}
\newcommand{\ddr}[1]{\left[\delta \langle r^2 \rangle ^2 \right]^{#1}}

\newcommand{\mdn}[2]{\overline{\delta \nu}_{#1}^{#2}}
\newcommand{\sdn}[2]{\widetilde{\delta \nu}_{#1}^{#2}}
\newcommand{\RomanNumeralCaps}[1]{\MakeUppercase{\romannumeral #1}}

\newcommand{\PP}{$^3P_0$}
\newcommand{\DD}{$^2D_{5/2}$}
\newcommand{\Dth}{$^2D_{3/2}$}
\newcommand{\Dtw}{$^1D_2$}
\newcommand{\JJ}{$4f^{13}5d6s^2 \: (J=2)$}
\newcommand{\FF}{$^2F_{7/2}$}
\newcommand{\Gtwo}{$G^{(2)}_i$}

\newcommand{\be}{\begin{equation}}
\newcommand{\ee}{\end{equation}}
\newcommand{\bea}{\begin{eqnarray}}
\newcommand{\eea}{\end{eqnarray}}

\begin{document}

\title{Orders-of-magnitude improvement in precision spectroscopy\\of an inner-shell orbital clock transition in neutral ytterbium}

\author{Taiki Ishiyama\,\orcidlink{0000-0002-5699-3713}}
\thanks{These authors contributed equally to this work. Corresponding author: ishiyama.taiki.88e@st.kyoto-u.ac.jp}
\affiliation{Department of Physics, Graduate School of Science, Kyoto University, Kyoto 606-8502, Japan}
\author{Koki Ono\,\orcidlink{0000-0002-1145-0424}}
\thanks{These authors contributed equally to this work. Corresponding author: ishiyama.taiki.88e@st.kyoto-u.ac.jp}
\affiliation{Department of Physics, Graduate School of Science, Kyoto University, Kyoto 606-8502, Japan}
\author{Hokuto Kawase\,\orcidlink{0009-0000-5075-9048}}
\affiliation{Department of Physics, Graduate School of Science, Kyoto University, Kyoto 606-8502, Japan}
\author{Tetsushi Takano\,\orcidlink{0000-0002-0406-4605}}
\affiliation{Department of Physics, Graduate School of Science, Kyoto University, Kyoto 606-8502, Japan}
\author{Reiji Asano\,\orcidlink{0009-0005-3861-6235}}
\affiliation{Department of Physics, Graduate School of Science, Kyoto University, Kyoto 606-8502, Japan}
\author{Ayaki Sunaga\,\orcidlink{0000-0002-3802-680X}}
\affiliation{ELTE, Eötvös Loránd University, Institute of Chemistry, Pázmány Péter sétány Budapest 1/A 1117, Hungary}
\author{Yasuhiro Yamamoto\,\orcidlink{0000-0003-1478-6043}}
\affiliation{Physics Division, National Center for Theoretical Sciences, National Taiwan University, Taipei 10617, Taiwan}
\affiliation{Accelerator Laboratory, High Energy Accelerator Research Organization (KEK), Tsukuba 305-0801, Japan}
\author{Minoru Tanaka,\orcidlink{0000-0001-8190-2863}}
\affiliation{Department of Physics, Graduate School of Science, The University of Osaka, Toyonaka 560-0043, Japan} 
\author{Yoshiro Takahashi\,\orcidlink{0000-0001-7607-7387}}
\affiliation{Department of Physics, Graduate School of Science, Kyoto University, Kyoto 606-8502, Japan}
\date{\today}

\begin{abstract}
An inner-shell orbital clock transition $\clock$ in neutral ytterbium atoms has attracted much attention as a new optical frequency standard as well as a highly sensitive probe for several new physics phenomena, such as ultralight dark matter, violation of local Lorentz invariance, and a new Yukawa potential between electrons and neutrons.
Here we demonstrate almost two-orders-of-magnitude improvement in precision spectroscopy over the previous reports on this transition, achieved by trapping atoms in a three-dimensional magic-wavelength optical lattice.
In particular, we successfully observe the coherent Rabi oscillation, the relaxation dynamics of the excited state and the interorbital Feshbach resonance.
To highlight the high precision of our spectroscopy, we carry out precise isotope shift measurements between five stable bosonic isotopes well below 10-Hz uncertainties, successfully setting bounds for a hypothetical boson mediating a force between electrons and neutrons.
These results open up the way for various new physics search experiments and a wide range of applications to quantum science with this clock transition.
\end{abstract}
\maketitle

Accuracy of optical clocks has been improved dramatically and is now reaching the $10^{-19}$ level~\cite{Aeppli2024-vr}, enabling the redefinition of the second~\cite{Dimarcq2024-uh} and various applications such as geodesy~\cite{Takano2016-vw, Lisdat2016-pi} and new physics searches~\cite{Safronova2018-xl}.
Furthermore, optical clock transitions are applied to quantum computation as optical clock qubits with non-destructive capability~\cite{Lis2023-go} and as intermediate states for Rydberg-state excitation in high-fidelity two-qubit gates~\cite{Madjarov2020-pu}, as well as the quantum simulation for a two-orbital physics such as a Kondo problem and as a high-resolution probe of quantum many-body systems~\cite{Schafer2020-bs}.

\begin{figure*}[!t]
\centering
\includegraphics[width = 0.9\linewidth]{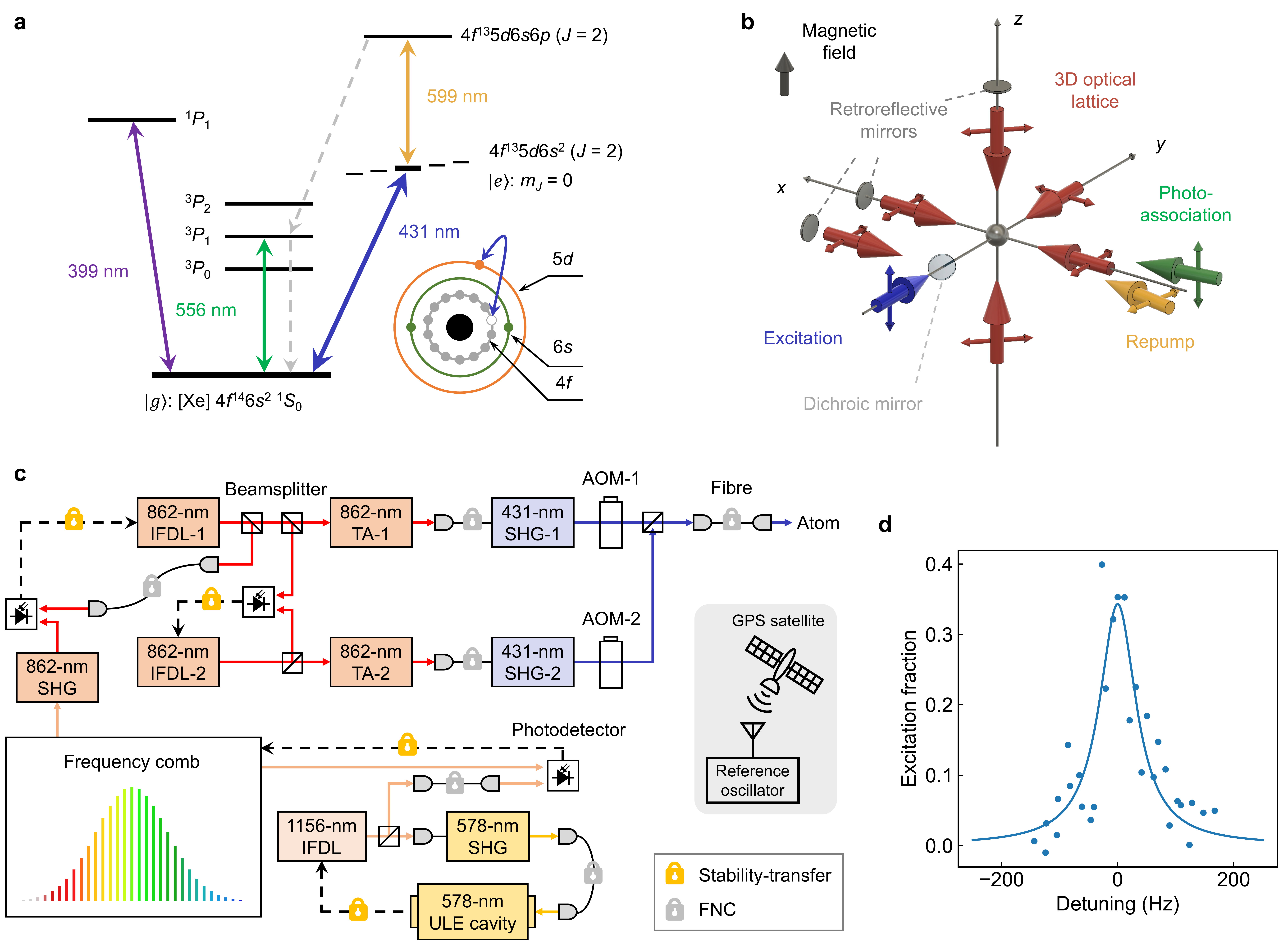}
\caption{\label{fig: setup}Experimental setup and precision spectroscopy. \textbf{a}, Energy diagram of neutral Yb atoms relevant to our experiments. \textbf{b}, Laser configuration for precision spectroscopy and IS measurements. Different directions of the magnetic field and polarizations of three optical lattice lasers are adopted for other measurements (Methods). \textbf{c}, Schematic diagram of our excitation laser system. Dashed arrows depict the locks of the stability transfer. The fibre-noise-cancellation (FNC) system~\cite{Ma1994-bh} is installed for all the optical fibres longer than three meters. All the radio-frequency sources used for the frequency stabilization are referenced to the global positioning system (GPS)-disciplined reference oscillator shown in the grey shaded region. Note that IFDL-1 is mainly used throughout this work, whereas IFDL-2 is used in IS measurements. TA, tapered amplifier. \textbf{d}, A typical spectrum of the $\ket{g} = \, ^1S_0 \leftrightarrow \ket{e} = \, 4f^{13}5d6s^2 \: (J=2, m_J = 0)$ transition in $^{174}$Yb. The blue solid line is a fitting curve by a Lorentzian function. The full-width at half-maximum and the maximum excitation fraction are 77(11) Hz and 0.34(3), respectively, where the values in parentheses are the $1\sigma$ fitting uncertainties. Note that $1\sigma$ uncertainties of excitation fractions from the absorption imaging are smaller than the symbol size.}
\end{figure*}

In 2018, a magnetic-quadrupole (M2) clock transition $\clock$ in neutral ytterbium (Yb) atoms at 431 nm (Fig.~\ref{fig: setup}a) was proposed in Refs.~\cite{Dzuba2018-kc, Safronova2018-ry}, which predicted the radiative lifetime of the excited state of 200 and 60~s, much longer than those of other metastable states.
Additionally, this clock transition exhibits high sensitivities to various new physics phenomena beyond the Standard Model (SM).
Specifically, it is sensitive to Lorentz violation in the electron-photon sector due to the large anisotropy of its electronic wave function~\cite{Shaniv2018-gc}, and to variations in the fine-structure constant (potentially induced by ultralight dark matter)~\cite{Dzuba2018-kc, Safronova2018-ry, Tang2023-ju}.
Furthermore, precise isotope shift (IS) measurement of this transition, combined with other clock transitions in Yb atoms~\cite{Ono2022-oy, Figueroa2022-cm} and Yb$^+$ ions~\cite{Counts2020-ca, Hur2022-ua, Door2025-jv}, allows us to study a nonlinearity of the King plot in the search for a hypothetical particle mediating a force between electrons and neutrons~\cite{Berengut2018-su, Mikami2017-oh, Tanaka2020-pb, Berengut2020-mx, Berengut2025-yy}.

The first experimental observation of this transition was reported in Ref.~\cite{Ishiyama2023-tv}, in which the magic wavelengths~\cite{Katori2003-hb} were determined to be around 797 nm and 834 nm towards precision spectroscopy.
Although this was followed by other studies on the $\clock$ ~\cite{Kawasaki2023-bx, Kawasaki2024-co, Wang2025-oi} and $^3P_0 \leftrightarrow \e$ transitions~\cite{Qiao2024-jf}, the narrowest linewidth of atomic spectra so far is the kilohertz-level~\cite{Wang2025-oi, Qiao2024-jf}, much broader than those used in modern optical clocks~\cite{Takano2016-vw}.

In this work, we demonstrate orders-of-magnitude improvement in precision spectroscopy of the magnetic insensitive $\ket{g} = \, ^1S_0 \: \leftrightarrow \: \ket{e} = 4f^{13}5d6s^2 \: (J=2, m_J=0)$ transition, where $m_J$ is the magnetic quantum number of the total angular momentum $J$.
We successfully observe an atomic spectrum with a linewidth of several tens of hertz, an improvement of about two orders of magnitude over the previous studies~\cite{Ishiyama2023-tv, Kawasaki2023-bx, Kawasaki2024-co, Wang2025-oi, Qiao2024-jf}.
In addition, we reveal important basic properties of this transition: the transition moment by observing the coherent Rabi oscillation, the intrinsic lifetime of the excited state and the interorbital Feshbach resonance.
To highlight the high precision of our spectroscopy, we measure ISs between five stable bosonic isotopes with total uncertainties well below 10~Hz.
Combining these data with ISs of other transitions~\cite{Counts2020-ca, Hur2022-ua, Door2025-jv, Ono2022-oy, Figueroa2022-cm}, we find a huge nonlinearity of $85 \sigma$ for a three-dimensional (3D) generalized King plot~\cite{Mikami2017-oh}, enabling constraints to the coupling constants of the new particle under reasonable assumptions, as well as implications for the Yb nucleus by the dual King plot approach~\cite{Yamamoto2023-zq}.
Our work opens up various new physics search experiments and wide applications to quantum science with the new clock transition.


\section*{\label{spectroscopy}Precision spectroscopy}

\begin{figure*}[ht]
\centering
\includegraphics[width = 0.9\linewidth]{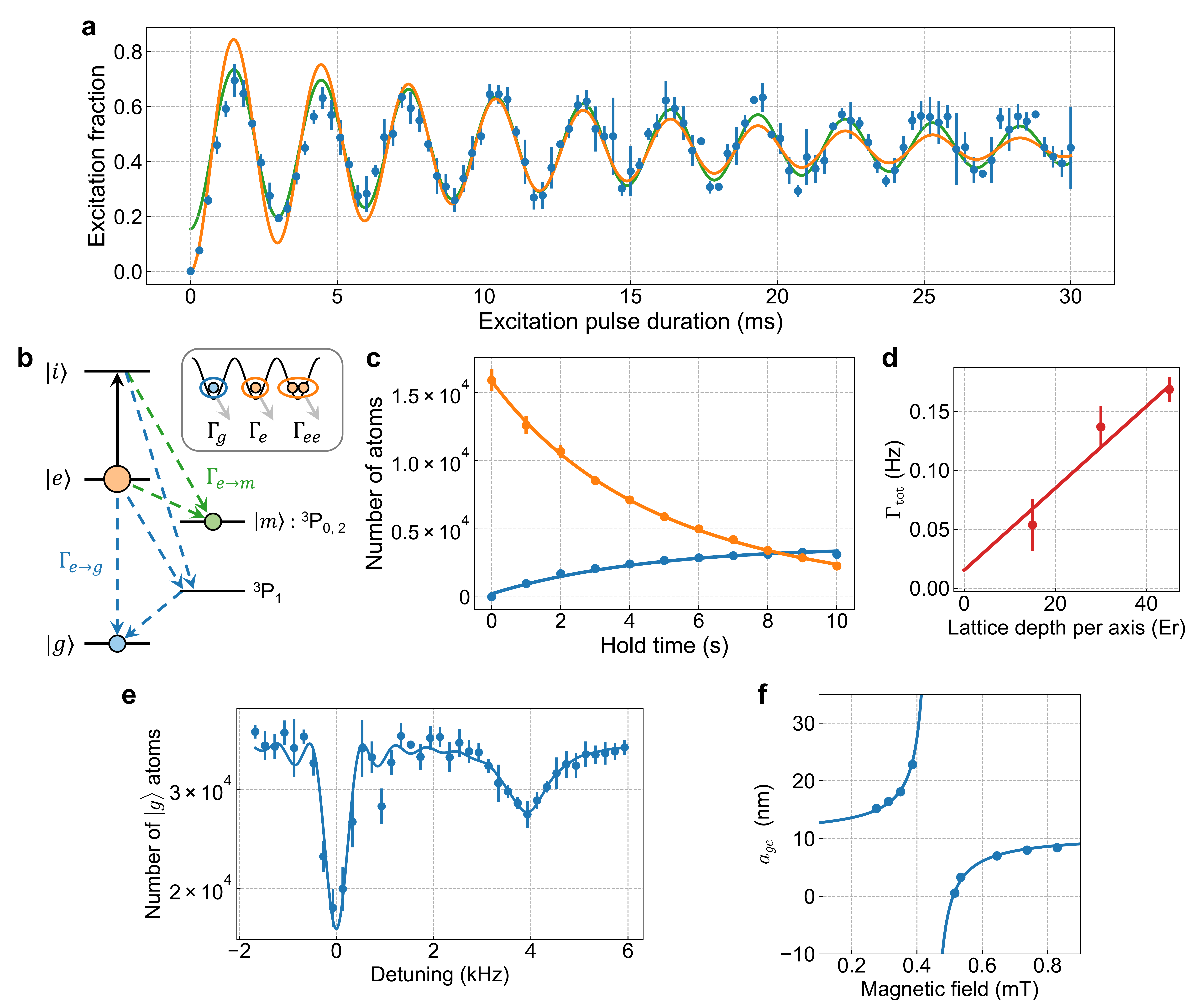}
\caption{\label{fig: basic} Basic properties of the new clock transition. \textbf{a}, Rabi oscillation. The blue points and error bars represent the average and standard deviation of three measurements, respectively. The orange solid line is a fitting curve by a sinusoidal function with an exponential damping. The Rabi frequency and the $1/e$ decay time are $2 \pi \times 335.8(7)$ Hz and $11.3(6)$ ms, respectively. When adding an offset to the fitting function, shown as the green solid line, the $1/e$ decay time is $20.1(1.9)$ ms with the Rabi frequency of $335.9(5)$~Hz. \textbf{b-d}, Lifetime measurement of the $\ket{e}$ state. \textbf{b}, Schematic of the loss mechanism of the $\ket{e}$ state. \textbf{c}, Relaxation dynamics of $\ket{g}$ (blue) and $\ket{e}$ (orange) atoms at the lattice depth of $45E_r$ per axis. The points and error bars represent the average and standard deviation of four measurements, respectively. The solid lines are the fitting curves by Eq.~(\ref{eq:lifetime_rate}).  \textbf{d}, Lattice depth dependence of the decay rate $\Gamma_\mathrm{tot}$. The data points and error bars represent the mean values and $1\sigma $ statistical uncertainties, respectively, derived from the fit. The solid line is a linear fit. \textbf{e-f}, Interorbital Feshbach resonance. \textbf{e}, Occupancy-resolved spectroscopy of $^{174}$Yb at the magnetic field of $0.276$~mT. A blue point and error bar are the average and standard deviation of four scans, respectively. The blue solid line is a fitting curve by a combination of a Rabi excitation spectrum for $n=1$ and a Lorentzian spectrum for $n=2$. \textbf{f}, Magnetic field dependence of $a_{ge}$. The blue solid line is a fitting curve by Eq.~(\ref{eq: Feshbach}), yielding $B_0 = 0.442(4)$~mT, $\Delta B = 0.068(4)$~mT and $a_{ge}^{\mathrm{BG}} = 10.6(5)$~nm. Numbers in parentheses are the $1\sigma$ total uncertainties, including fitting and systematic uncertainties, such as the lattice depth and magnetic field calibration.}
\end{figure*}

Figure~\ref{fig: setup}b shows the laser configuration for atom cooling, trapping and probing.
Ultracold Yb atoms with an atomic temperature of 300 nK are prepared in a 3D isotropic optical lattice at the depth of $30 E_r$ per axis.
The frequency of the optical lattice is set to $376.054$~THz, which is the magic condition for $^{174}$Yb when the laser polarization is perpendicular to the quantization axis (Supplementary Information, section \RomanNumeralCaps{1}).
Here $E_r = h \times 1.8$~kHz is the recoil energy of $^{174}$Yb at 797.2 nm, and $h$ is the Planck constant.
Then, we shine a photo-association laser red-detuned from the $^1S_0$ $\leftrightarrow$ $^3P_1$ resonance to remove multiply occupied sites.

Subsequently, a 431-nm excitation laser with a peak intensity of 1.8 W cm$^{-2}$ is irradiated for $200$~ms along the $y$ axis lattice.
Figure~\ref{fig: setup}c illustrates our excitation laser system.
An interference-filter-stabilized external-cavity diode laser (IFDL) at a fundamental wavelength of 862 nm is stabilized to an optical cavity made of ultralow-expansion (ULE) glass via an optical frequency comb using the stability transfer method~\cite{Yamaguchi2012-zf}.
An acousto-optic modulator (AOM-1) is used to fine-tune 431-nm laser frequencies to perform the spectroscopy.
A magnetic field of $14.6$~mT is applied along the $z$ axis to set the quantization axis and induce the magnetic-field-induced electric-dipole (E1) transition~\cite{Taichenachev2006-cu, Ishiyama2023-tv}.
The number of $\ket{g}$ atoms is measured by the absorption imaging with the $\g \leftrightarrow \: ^1P_1$ transition, whereas that of $\ket{e}$ atoms is determined in combination with repumping using the $\e \leftrightarrow 4f^{13}5d6s6p \: (J=2)$ transition.

Figure~\ref{fig: setup}d shows a typical atomic spectrum of the $\gtoe$ transition in $^{174}$Yb.
The full-width at half-maximum is typically about 80~Hz, two-orders-of-magnitude narrower than previous studies~\cite{Ishiyama2023-tv, Kawasaki2023-bx, Kawasaki2024-co, Wang2025-oi, Qiao2024-jf}.
The linewidth may be limited by the instability of the ULE cavity and the inhomogeneity of the probe light shift, an a.c. Stark shift induced by the excitation laser itself.
The demonstration of precision spectroscopy allows us to investigate fundamental properties of this new clock transition.

\section*{\label{Rabi}Coherent control of optical clock transition}
First, we present the coherent excitation of the clock transition, essential for the application to quantum computation~\cite{Lis2023-go, Madjarov2020-pu} and quantum simulation\cite{Schafer2020-bs}.
It is also important to experimentally determine the transition rate to test the atomic theoretical calculation~\cite{Dzuba2018-kc}.

Figure~\ref{fig: basic}a shows the observed Rabi oscillation of $^{174}$Yb with the M2 transition mechanism.
The observed decay time of $11.3(6)$ ms or $20.1(1.9)$ ms, which depends on a fitting function, is consistent with the typical linewidth of atomic spectra.
The M2 transition moment is determined as $\bra{J_e} | Q^{\mathrm{mg}}_{2} | \ket{J_g} = 1.69(9) \times 10^{-33} \ \mathrm{A} \, \mathrm{m}^3$, almost in agreement with the theoretical value of $1.3 \times 10^{-33} \ \mathrm{A} \, \mathrm{m}^3$~\cite{Dzuba2018-kc} (Methods).
We also determine the relative strength of the magnetic-field-induced E1 transition~\cite{Lange2021-wz, Klusener2024-yh} as $2\pi \times 59.8(1.8)$~Hz by observing the coherent Rabi oscillation (Supplementary Information, section \RomanNumeralCaps{2}).

\section*{\label{lifetime}Intrinsic lifetime of the \texorpdfstring{$\e$}{e} state}
The lifetime of the excited state of a clock transition is crucial since it is directly related to the clock's quality factor.
The theoretical calculations estimated the lifetime of the $\e$ state to be 200~s~\cite{Dzuba2018-kc}, 60~s~\cite{Safronova2018-ry} and 190~s~\cite{Tang2023-ju}, with the dominant decay channel of the magnetic-dipole transition to the short-lived $^3P_1$ state.
In Ref.~\cite{Ishiyama2023-tv}, the trap lifetime of the excited state was reported as $1.9(1)$~s, much shorter than these predictions, possibly due to photon scattering by a $797$-nm lattice laser near-resonant to the $4f^{13}5d6s^2 \: (J=2) \leftrightarrow 4f^{13}6s^26p_{3/2} \: (J=3)$ transition at 792 nm.
Here we present measurements and analysis to obtain more accurate information on the intrinsic lifetime of the excited state.

The time dependence of the atom number in the $\ket{g}$ and $\ket{e}$ states is formulated as (Fig.~\ref{fig: basic}b)
\begin{equation}
\label{eq:lifetime_rate}
\begin{cases}
    \dot{N}_g = \Gamma_{e \rightarrow g} N_e -\Gamma_g N_g \\
    \dot{N}_e = -\Gamma_{\mathrm{tot}} N_e - \Gamma_{ee} N_e^2.
\end{cases}
\end{equation}
Here $N_{g(e)}$ is the number of $\ket{g(e)}$ atoms.
$\Gamma_\mathrm{tot} \equiv \Gamma_{e \rightarrow g} + \Gamma_{e \rightarrow m} + \Gamma_e$ represents the total decay rate from $\ket{e}$, where $\Gamma_{e \rightarrow g}$ is the decay rate to $\ket{g}$ and $\Gamma_{e \rightarrow m}$ is that to other metastable states $\ket{m}$ such as $^3P_0$ and $^3P_2$.
These decays could be caused by several mechanisms, such as the decay from an intermediate state $\ket{i}$ to which $\ket{e}$ atoms are excited by the photon scattering by lattice lasers and/or blackbody radiation (BBR)~\cite{Yasuda2004-hy, Walhout1995-pl}, as well as the intrinsic decay channel of the $\ket{e}$ state.
Additionally, several loss mechanisms from the optical lattice are considered; $\Gamma_{g(e)}$ denotes the loss rate for $\ket{g(e)}$ atoms presumably due to background-gas collisions and $\Gamma_{ee}$ represents a two-body loss rate due to an inelastic collision between two $\ket{e}$ atoms.

Figure~\ref{fig: basic}c shows the hold-time dependence of $N_g$ and $N_e$ in the 3D optical lattice at a depth of $45E_r$ per axis.
By fitting the data with Eq.~(\ref{eq:lifetime_rate}), we obtain $\Gamma_{e \rightarrow g}=0.050(2)$~Hz, $\Gamma_\mathrm{tot}=0.169(10)$~Hz and $\Gamma_{ee}=2.9(1.2)\times10^{-6}$~Hz.
Figure~\ref{fig: basic}d depicts the lattice depth dependence of $\Gamma_\mathrm{tot}$, indicating that the photon scattering by the optical lattice strongly contributes to the decay.
By a linear fit, the decay rate of the $\ket{e}$ state without the photon scattering effect is determined as $\Gamma_\mathrm{tot} = 0.02(4)$~Hz, corresponding to a lifetime of $\tau = 1/\Gamma_\mathrm{tot} = 66$~s with a lower bound of $20$~s (Methods).
The obtained lifetime is somewhat shorter than the theoretical values~\cite{Dzuba2018-kc, Tang2023-ju}, possibly due to the trap loss of the excited state denoted as $\Gamma_e$, and/or BBR-induced quenching~\cite{Walhout1995-pl, Yasuda2004-hy}, in which $\ket{e}$ is excited to upper states $\ket{i}$ by absorbing BBR photons and decays to others.
Our calculation shows the BBR quenching rate of the $\ket{e}$ state is about 0.060~Hz, suggesting that the measured decay rate is limited by the BBR quenching and the intrinsic lifetime should be even longer (Methods and Extended Data Fig.~\ref{fig:lifetime_BBR}).

\section*{\label{Feshbach}Interorbital Feshbach resonance}
The Feshbach resonance is a key tool in quantum simulation with ultracold atomic systems, by which an interatomic interaction can be controlled~\cite{Chin2010-zv}.
Here we report the observation of a magnetic Feshbach resonance between $\ket{g}$ and $\ket{e}$ atoms in $^{174}$Yb, possibly induced by the anisotropic interaction~\cite{Kato2013-jz}.

In occupancy-resolved spectroscopy in a deep 3D optical lattice, an interorbital interaction characterized by the $s$-wave scattering length $a_{ge}$ appears as a frequency shift between the $n=1$ and $n=2$ sites~\cite{Kato2013-jz}, where $n$ denotes the occupation number in a lattice site.
Figure~\ref{fig: basic}e depicts an occupancy-resolved spectrum under a magnetic field of $0.276$~mT.
The left signal corresponds to $n=1$ sites and the right, to $n=2$, confirmed by applying the photo-association laser that induces loss only for $n=2$ sites.
Figure~\ref{fig: basic}f shows the magnetic field dependence of $a_{ge}$.
By fitting the data with a function of~\cite{Chin2010-zv} 
\begin{equation}\label{eq: Feshbach}
    a_{ge}(B) = a_{ge}^{\mathrm{BG}} \left(1 - \frac{\Delta B}{B - B_0} \right),
\end{equation}
we successfully identify the Feshbach resonance at $B_0 = 0.442(4)$~mT (Methods).

\section*{\label{IS}Isotope shift measurement}

\begin{figure*}[!t]
\centering
\includegraphics[width = 0.95\linewidth]{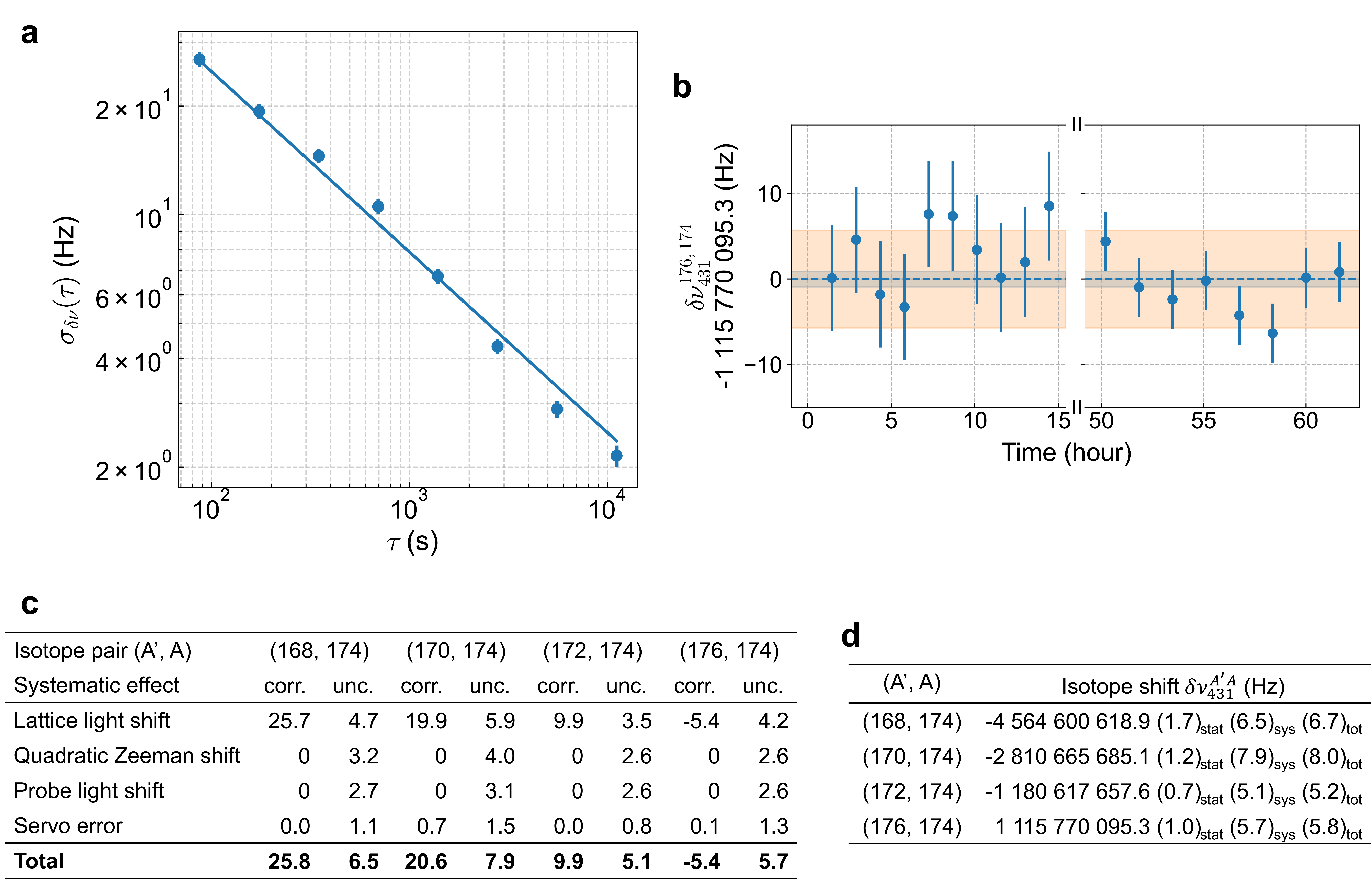}
\caption{IS measurements. \textbf{a}, Overlapping Allan deviation $\sigma_{\delta \nu} (\tau)$ representing stability of $\dn{431}{176, 174}$ measurements. The error bars indicate $1\sigma$ standard errors. The blue solid line is a fitting line assuming a white frequency noise, yielding $\sigma_{\delta \nu}(\tau) = 249(8)~\mbox{Hz} / \sqrt{\tau}$. \textbf{b}, Time traces of $\dn{431}{176, 174}$. A data point and error bar show the mean value and the overlapping Allan deviation of each segment, respectively. Blue (orange) shaded regions represent the $1\sigma$ statistical (systematic) uncertainties. Note that the stability shown in \textbf{a} corresponds to the latter half of \textbf{b}. \textbf{c}, Systematic corrections and uncertainties for $\dn{431}{A'A}$ in Hz. AOM chirp and BBR shift are not shown here since they are negligibly small. \textbf{d}, Summary of measured ISs. Statistical, systematic and total $1\sigma$ uncertainties are shown in the parentheses.}
\label{fig: IS}
\end{figure*}

To highlight the high precision of our spectroscopy, we measure ISs between five stable bosonic isotopes, which directly leads to the test of a hypothetical Yukawa potential between electrons and neutrons.
Hereafter, an IS is denoted as $\dn{i}{A'A} = \nu_i^{A'} - \nu_i^{A}$, where $A = 168, 170, 172, 174, 176$ is a label of an isotope $^A$Yb, $i$ denotes a transition with a wavelength in units of nanometres ($i=431$ for this transition) and $\nu_i^{A}$ is a transition frequency.

To achieve good stability and accuracy, we adopt an interleaved clock operation between two isotopes~\cite{Ono2022-oy}, where we can mitigate most of the systematic effects common to isotopes, such as the resonance frequency drift of the ULE cavity and intensity fluctuation of optical lattice lasers.
Figure~\ref{fig: IS}a shows the overlapping Allan deviation $\sigma_{\delta \nu}(\tau)$ of $\dn{431}{176, 174}$ with respect to the averaging time $\tau$.
The blue solid line is a fitting curve assuming a white frequency noise, indicating that serious long-term perturbation is not observed in the measurement.
For each isotope pair, we perform two measurements on different days, each divided into about ten segments.
Figure~\ref{fig: IS}b displays the time traces of $\dn{431}{176, 174}$, illustrating that the two measurements performed on two distinct days are consistent within the systematic uncertainty.
Figure~\ref{fig: IS}c shows all systematic corrections and uncertainties of our IS measurements.
For lattice light shifts that give the largest systematic corrections, we confirm their validity by measuring the ISs of the $4f^{13}5d6s^2 \: (J=2) \leftrightarrow 4f^{13}6s^26p_{3/2} \: (J=3)$ transition at 792 nm, for which the lattice laser is near-resonant (Supplementary Information, section \RomanNumeralCaps{3}).

Figure~\ref{fig: IS}d summarizes the measured ISs with the $1\sigma$ statistical, systematic and total uncertainties.
As a result, we successfully achieve the total uncertainties of less than $8.0$~Hz, which is an improvement of four orders of magnitude~\cite{Kawasaki2024-co} and comparable with the most precisely measured ISs of other transitions in Yb~\cite{Ono2022-oy, Door2025-jv} (Methods and Extended Data Figs.~\ref{fig: IS_time_trace}-\ref{fig: IS_PLS} provide details of the measurement and analysis).

\section*{\label{KP}King plot analysis of isotope-shift data}
The obtained precision IS data are utilized for King plot analysis to set constraints for new physics as well as nuclear physics.
In general, an IS is written as~\cite{Berengut2025-yy}
\begin{equation}\label{eq: IS}
\begin{split}
    \dn{i}{A'A}
    = \ &K_i \w{A'A} + F_i \dr{A'A} + H_i \delta \eta^{A'A} \\
      &+ \alpha_{\mathrm{NP}} X_i (A'-A).
\end{split}
\end{equation}
The first two terms stem from the SM, namely, the leading-order mass and field shifts, respectively.
The third is a higher-order SM term, and in the case of Yb, it corresponds to a higher-order field shift such as the next-leading-order Seltzer moment $G^{(4)}_i \drr{A'A}$ and/or the quadratic field shift (QFS) $G^{(2)}_i \ddr{A'A}$.
Each term is factorized to the transition-dependent electronic factor and the isotope-dependent nuclear factor.
$\w{A'A} = m_{172}/m_{A'} - m_{172}/m_A$ is obtained by the nuclear mass ratios measured at the level of $10^{-12}$~\cite{Door2025-jv}, where $m_{A}$ is the nuclear mass of isotope $A$.
$\delta \langle r^n \rangle ^{A'A} = \langle r^n \rangle ^{A'} - \langle r^n \rangle ^A$ is the difference of the $n$th nuclear charge moment, and $\ddr{A'A}$ is defined as $(\dr{A'A_0})^2 - (\dr{AA_0})^2$ with the reference isotope $A_0$.
The final term is the particle shift (PS) induced by a new Yukawa potential between electrons and neutrons:
\begin{equation}\label{eq:Yukawa}
    V(r) = \left( -1 \right)^{1+s_\phi} y_e y_n \frac{e^{-m_{\phi}cr/\hbar}}{4\pi r}.
\end{equation}
Here $s_\phi$ and $m_{\phi}$ are the spin and mass of the new particle, respectively.
$y_{e(n)}$ represents the coupling constant between the new particle and an electron (neutron), whereas $\alpha_{\mathrm{NP}}$ is defined as $(-1)^{1+s_\phi} y_e y_n / (4 \pi \hbar c)$.
The electronic factors $X_i$ are given in Extended Data Fig.~\ref{fig: X} (Supplementary Information, section \RomanNumeralCaps{4}).
In particular, the $431$-nm transition possesses a larger absolute value of $X_i$ than those of the other transitions except for the 467-nm transition (see the lighter mass region $m_\phi \lesssim 1$~keV in Extended Data Fig.~\ref{fig: X}).
Note that the $m_\phi$ dependence of the 431-nm transition is similar to that of the 467-nm transition since the $4f$ electron is excited in both transitions.

\subsection*{\label{sec: original KP}King plot and constraints on new physics}

\begin{figure*}[!tb]
\centering
\includegraphics[width = 0.9\linewidth]{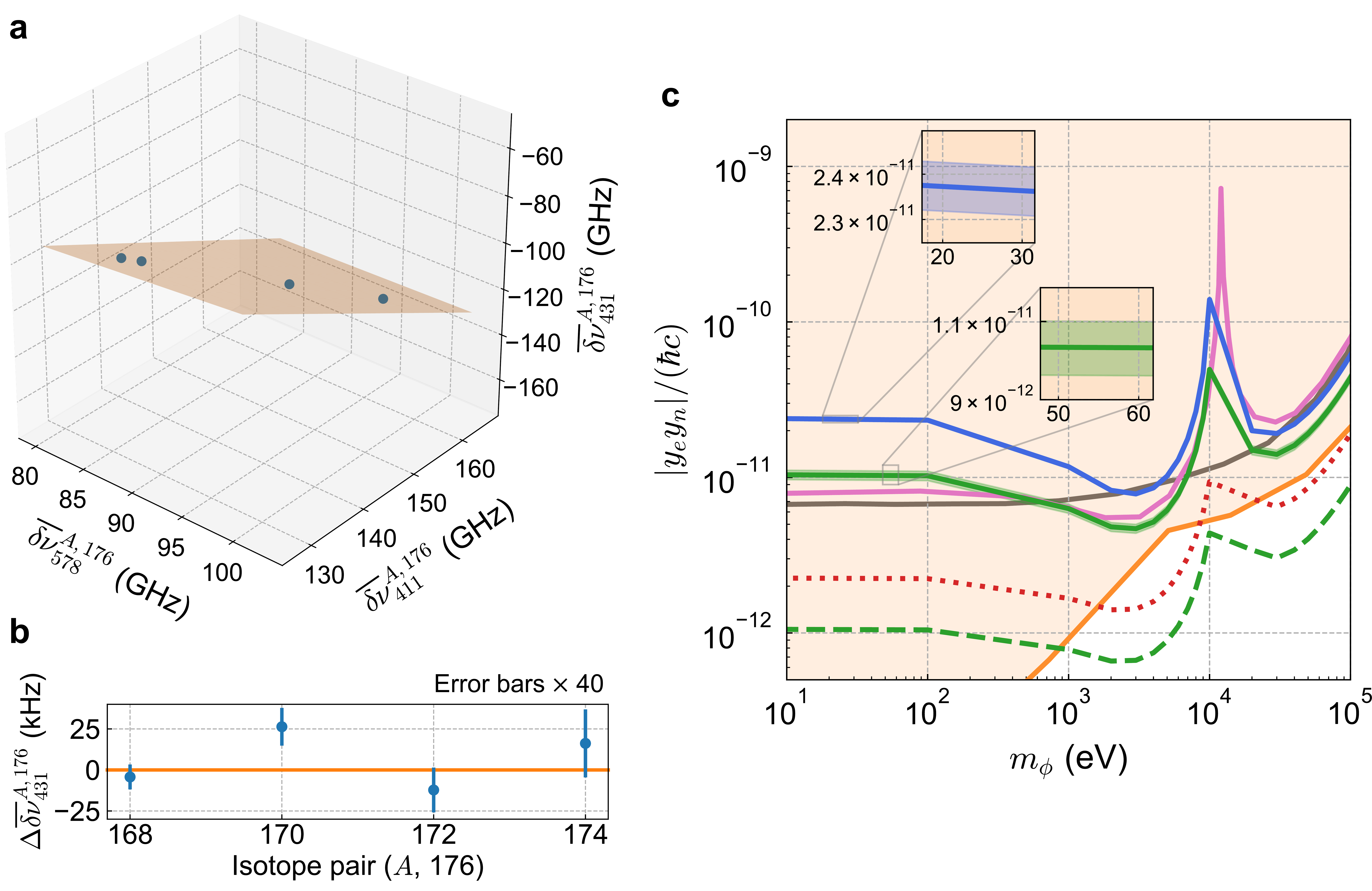}
\caption{King plot analysis. 
\textbf{a}, 3D generalized King plot. Three axes represent modified ISs of $578$-nm~\cite{Ono2022-oy}, $411$-nm~\cite{Door2025-jv} and $431$-nm (this work) transitions with the nuclear mass ratio measurement~\cite{Door2025-jv}. The orange plane is a fit to four data points, corresponding to the 3D generalized King relation. 
\textbf{b}, Residuals from the best-fit plane along the axis of the $431$-nm transition. The error bars represent $1\sigma$ statistical uncertainties of modified ISs of the $431$-nm transition, which are multiplied by 40 for visualization.
\textbf{c}, Product of couplings $|y_e y_n|$ of a new boson as a function of the mass $m_\phi$. 
The blue line and shaded area are the central value and the 95\% confidence interval, respectively, under the assumption that the PS is the origin of the nonlinearity. 
The green solid line and shaded area represent the favoured region with QFS-subtracted ISs.
The green dashed line is the upper bound with QFS-subtracted ISs when selecting $G^{(2)}_i$ ratio values to minimize the nonlinearity.
The red dotted line represents the prospected upper bound with an unstable isotope of $^{166}$Yb.
All shaded bands and upper bounds are evaluated at the 95\% confidence level.
For comparison, the constraint on $y_e$ from a $(g-2)_e$ measurement~\cite{Fan2023-cz}, times the constraint on $y_n$ from neutron scattering measurements~\cite{Leeb1992-vi, Pokotilovski2006-ct, Nesvizhevsky2008-mf} is shown as an orange solid line, and the orange shaded region is the corresponding excluded area.
The pink solid line shows the favoured region from Yb King plot with the 578, 411, 467 transitions~\cite{Door2025-jv}.
The upper bound from the King plot analysis of calcium ions~\cite{wilzewski2024nonlinearcalciumkingplot} is displayed as a grey solid line.}
\label{fig: KP}
\end{figure*}

Considering only the two leading-order terms in Eq.~(\ref{eq: IS}), we obtain a two-dimensional linear relation between two transitions $i$ and $j$ called the two-dimensional King linearity~\cite{King1963-sb}:
\begin{equation}\label{eq: original 2D KP}
     \mdn{j}{A'A} = K_{ij} + F_{ij} \mdn{i}{A'A},
\end{equation}
where $F_{ij} = F_j / F_i$ and $K_{ij} = K_j - F_{ij} K_i$ are coefficients dependent only on transitions, and $\mdn{i}{A'A} = \dn{i}{A'A} / \w{A'A}$ is the modified IS.
This relationship can be generalized to higher dimensions by incorporating additional transitions~\cite{Mikami2017-oh}.

By combining the measured ISs with those of the $^1S_0 \leftrightarrow \, ^3P_0$ transition ($i=578$) in Yb~\cite{Ono2022-oy} as well as the $^2S_{1/2} \leftrightarrow \, ^2D_{5/2}$ transition ($i=411$) in Yb$^+$~\cite{Door2025-jv}, we can construct the 3D generalized King plot (Fig.~\ref{fig: KP}a).
Although the data look almost aligned in a plane, Fig.~\ref{fig: KP}b shows that the 3D generalized King linearity is clearly violated, corresponding to $\chi^2 = 7.2 \times 10^3$ and $85\sigma$ significance (Supplementary Information, section \RomanNumeralCaps{5}).
This means that the data cannot be explained by a single higher-order effect such as the QFS, the next-leading-order Seltzer moment, or the PS, but at least two distinct higher-order terms are involved in the data. 
To extract the information on the new particle, therefore, we need to assume the origin of the observed nonlinearity.

By naively assuming that the PS is the origin of this nonlinearity, we obtain the favoured region for the product of the coupling constants $|y_e y_n|$, as shown by the blue solid line and shaded area in Fig.~\ref{fig: KP}c  (Methods).
The obtained region conflicts with the best terrestrial bound~\cite{Fan2023-cz, Leeb1992-vi, Pokotilovski2006-ct, Nesvizhevsky2008-mf} (shown as the orange solid line and shaded area), suggesting that the observed nonlinearity originates not only from the PS but also from the higher-order SM terms.
Using the $^2S_{1/2} \leftrightarrow \, ^2F_{7/2}$ transition ($i=467$) in Yb$^+$~\cite{Door2025-jv} instead of the 411-nm transition provides another 3D King plot. 
Simultaneous analysis of these two independent linear relations results in a tighter constraint on PS (Supplementary Information, section \RomanNumeralCaps{5}).

To obtain information on the new particle in the presence of two higher-order SM terms, the generalized King plot approach requires IS data for at least four distinct transitions with five isotope pairs.
Although the precision IS data are available for four optical transitions, the number of isotope pairs is limited to four, rendering us to consider a subtracted IS approach based on a theory input.
Here we assume that QFS is one of the two higher-order terms contributing to the observed nonlinearity.
The QFS can be subtracted from the measured ISs by using the reference transition $i_0$ and theoretically calculated ratios of $G_i^{(2)}$ (Methods and Extended Data Table~\ref{tab: electronic factor}):
\begin{equation} \label{eq: QFS-subtracted IS}
    \sdn{i}{A'A} = \dn{i}{A'A} - \frac{G^{(2)}_i}{G^{(2)}_{i_0}} \dn{i_0}{A'A}.
\end{equation}
Here we synthesize QFS-subtracted ISs $\sdn{i}{A'A} \ (i=431, 411, 467)$ with $i_0=578$ and construct the generalized 3D King plot, resulting in the nonlinearity down to $\chi^2 = 1.0 \times 10^3$ and $32 \sigma$ significance (Supplementary Information, section \RomanNumeralCaps{5}).
Assuming that this nonlinearity originates from the PS, the favoured region shown as the green solid line and shaded area in Fig.~\ref{fig: KP}c.
Alternatively, we can select $G^{(2)}_i$ ratios to minimize the nonlinearity and set the stringent upper bound, shown as a green dashed line in Fig.~\ref{fig: KP}c (Methods).

Finally, we discuss the prospect of the Yb King plot approach by additionally using a radioactive bosonic isotope $^{166}$Yb with a half-life of $2.4$ days.
Here we assume that $\dn{i}{166, 176} \ (i = 578, 431, 411, 467)$ and $\w{166, 172}$ are determined with the same precision as that for stable isotope pairs, and the four-dimensional (4D) King linearity holds.
Then, the upper bound is obtained as the red dotted line in Fig.~\ref{fig: KP}c, suggesting that the Yb King plot approach might improve the best terrestrial bound without any additional theoretical input, especially in the mass range of around $10^3$--$10^5$ eV (Methods).

\subsection*{\label{dual KP}Dual King plot and constraints on nuclear factors}
\begin{figure*}[!tb]
\centering
\includegraphics[width = 0.9\linewidth]{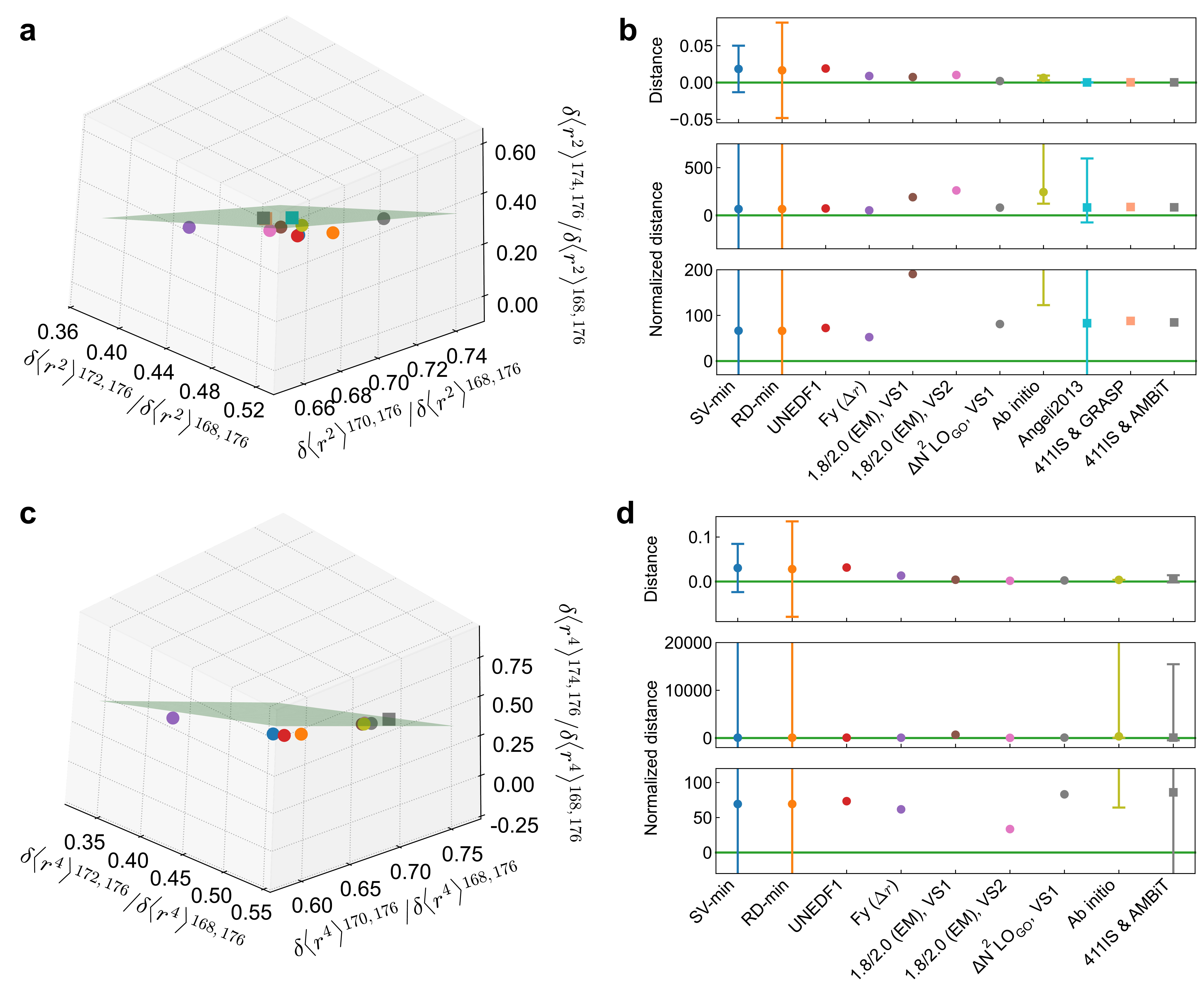}
\caption{Constraints on nuclear factors by the dual King plot analysis. \textbf{a}, A constraint on $\dr{}$. The green plane represents the linear relation Eq.~(\ref{subeq: dual dr}) with $\alpha_m$ determined by the dual King plot analysis. Circles are theoretical values based on nuclear calculation, while squares are values obtained from the theoretically calculated electronic factors and experimentally measured ISs.
\textbf{b}, Deviations of the points from the constraint plane. The top figure shows the distance of each point from the best plane. The middle displays the distance normalized by the $1 \sigma$ uncertainty of the plane, while the bottom is its magnified view. 
[SV-min; RD-min; UNEDF1; Fy($\Delta$r); 411IS\&GRASP; 411IS\&AMBiT]:~\cite{Hur2022-ua}, 
[1.8/2.0 (EM), VS1 and VS2; $\Delta \mathrm{N}^2 \mathrm{LO}_{\mathrm{GO}}$, VS1; Ab initio]:~\cite{Door2025-jv},
[Angeli2013]:~\cite{Angeli2013-pc}.
`Ab initio' is defined as the weighted average of `1.8/2.0 (EM), VS1' and `$\Delta \mathrm{N}^2 \mathrm{LO}_{\mathrm{GO}}$, VS1'. 
\textbf{c, d}, A constraint on $\drr{}$.
[SV-min; RD-min; UNEDF1; Fy($\Delta$r)]:~\cite{Hur2022-ua}, 
[1.8/2.0 (EM), VS1 and VS2; $\Delta \mathrm{N}^2 \mathrm{LO}_{\mathrm{GO}}$, VS1; Ab initio; 411IS\&AMBiT]:~\cite{Door2025-jv}.
Note that the error bars are associated with the $1\sigma$ uncertainties given in their references.
}
\label{fig: dual KP}
\end{figure*}

Next, we present the dual King plot analysis~\cite{Yamamoto2023-zq}.
In contrast to the original King plot, the role of isotope pairs and transitions are exchanged in the dual King plot; the IS for each transition is plotted in the space spanned by the isotope pairs.
These points satisfy the linear relationship called the dual King relation, whose coefficients provide stringent constraints on the nuclear factors.

Here we suppose that the contribution of the PS is negligibly small within the current measurement precision, and the IS consists of the two leading-order SM terms and the two higher-order field shift terms.
Then, the IS and the nuclear factors are shown to satisfy the following equations (Methods provides the derivation):
\begin{align}
    &\sum_{m} \alpha_m \dn{i}{A_mA_0} = G^{(2)}_i, \label{subeq: dual IS} \\
    &\sum_{m} \alpha_m \w{A_mA_0} = 0, \label{subeq: dual mu} \\
    &\sum_{m} \alpha_m \dr{A_mA_0} = 0, \label{subeq: dual dr} \\
    &\sum_{m} \alpha_m \drr{A_mA_0} = 0, \label{subeq: dual drr} \\
    &\sum_{m} \alpha_m \ddr{A_mA_0} = 1, \label{subeq: dual ddr}
\end{align}
where $m=1,2,3,4$ correspond to four isotope pairs consisting of four isotopes $A_m = 168, 170, 172, 174$, respectively, with respect to the reference isotope $A_0 = 176$.
By determining $\alpha_m$ by a fit with Eqs.~(\ref{subeq: dual IS}) and (\ref{subeq: dual mu}), constraints on $\dr{A_mA_0}$ and $\drr{A_mA_0}$ are acquired from Eqs.~(\ref{subeq: dual dr}) and (\ref{subeq: dual drr}).

Figure~\ref{fig: dual KP} shows constraints for $\dr{}$ and $\drr{}$ obtained by using the ISs of the 578-nm, 431-nm, 411-nm and 467-nm transitions as well as nuclear mass ratios, which are compared with theoretical and experimental values~\cite{Hur2022-ua, Door2025-jv, Angeli2013-pc} (Methods).
As shown in Fig.~\ref{fig: dual KP}, the dual King plot analysis gives us much more stringent constraints for nuclear factors than the previous experiments~\cite{Angeli2013-pc, Hur2022-ua}, labelled as Angeli2013 and 411IS\&AMBiT in Fig.~\ref{fig: dual KP}b,d.
We note that the obtained constraints can be used as a benchmark for nuclear theories.

\section*{\label{Discussion}Discussion}
In this work, we report the orders-of-magnitude improvement in precision spectroscopy of the $\clock$ transition of Yb atoms.
By this development, we reveal important basic properties of this inner-orbital clock transition: the transition moment, the intrinsic lifetime of the excited state and the interorbital Feshbach resonance.
Moreover, we measure ISs with uncertainties of less than 10~Hz, which are used to set constraints on new physics and nuclear parameters with the King plot approach.

We show that the intrinsic lifetime of the excited state is longer than that of other metastable states calculated in Ref.~\cite{Dzuba2018-kc}. 
This indicates that the inner-shell clock transition has the potential to improve the performance of the optical clock beyond that of the $^1S_0 \leftrightarrow \, ^3P_0$ transition of $^{171}$Yb, which is adopted as the secondary representation of the second.
Also, our successful observation of the Feshbach resonance opens the door to the quantum simulation of the two-orbital Bose-Hubbard model with controllable interaction.
Especially, it is interesting to search Feshbach resonances for the fermionic isotopes $^{171}$Yb and $^{173}$Yb towards studies of the crossover between a Bose-Einstein condensate and a Bardeen-Cooper-Schriefer superfluid~\cite{Chin2010-zv}.
Our King plot analysis of the observed precision IS measurements reveals a huge nonlinearity of $85\sigma$, enabling constraints on the coupling constants of the new particle under reasonable assumptions, as well as implications for the Yb nucleus by the dual King plot approach.
To further improve the constraint on new physics, IS measurements and nuclear mass measurement including the unstable $^{166}$Yb~\cite{Saito2019-qf} are highly desired.
Finally, new physics search experiments using the inner-orbital clock transition, such as ultralight dark matter~\cite{Dzuba2018-kc, Safronova2018-ry, Tang2023-ju} and Lorentz violation~\cite{Shaniv2018-gc}, are fairly promising. 
Our work opens up wide applications of this new clock transition.

\section*{Acknowledgments}
We thank Amar Vutha,  Tomoya Naito and Naoya Ozawa for useful discussions.
This work was supported by
the Grant-in-Aid for Scientific Research of JSPS (No. JP17H06138, No. JP18H05405, No. JP18H05228, No. JP21H01014, No. JP22K20356, No. 24K16995, No. 24KJ1347, No. 24K07018), 
JST PRESTO (No. JP-MJPR23F5), 
JST CREST (Nos. JPMJCR1673 and JP-MJCR23I3), 
MEXT Quantum Leap Flagship Program (MEXT Q-LEAP) Grant No. JPMXS0118069021, 
JST Moon-shot R\&D (Grant No. JPMJMS2268 and JP-MJMS2269), 
and JST ASPIRE (No. JPMJAP24C2).
KO acknowledges support from Graduate School of Science, Kyoto University under Ginpu Fund.
AS acknowledges financial support from the JSPS KAKENHI (Grant No. 21K14643).
The work of YY was supported by the National Science and Technology Council, the Ministry of Education (Higher Education Sprout Project NTU-112L104022), and the National Center for Theoretical Sciences of Taiwan.
We used the supercomputer of ACCMS, Kyoto University (Service Course and Collaborative Research Project for Enhancing Performance of Programming), and the computer resource offered under the category of General Projects by the Research Institute for Information Technology, Kyushu University. 

\section*{Author contributions}
TI, KO, HK, TT and RA carried out experiments.
TI, KO, YY and MT implemented the data analysis.
AS carried out the atomic theoretical calculation.
YT supervised the whole project.
All the authors contributed to the writing of the manuscript.

\section*{Competing interests}
The authors declare no competing interests.

\section*{Data availability}
The data supporting the results presented in this paper are provided with this paper as Source data files.


\clearpage
\section*{Methods}

\subsection*{Experimental setups}
Ultracold Yb atoms are prepared in the same manner as in our previous works~\cite{Ono2022-oy, Ishiyama2023-tv}.
For the optical lattice, we use a titanium-sapphire laser as the $x$ axis lattice laser.
It is also used as a seed light for two TAs whose output light beams are utilized for the $y$- and $z$ axis lattices.
Note that we make use of volume Bragg gratings to eliminate amplified spontaneous emissions.
The lattice depth is calibrated using a pulsed lattice technique~\cite{Denschlag2002-hb}, and the laser power is stabilized with AOMs.

For frequency stabilization of the clock excitation lasers at 431 nm, we utilize a stability transfer method via an optical frequency comb~\cite{Yamaguchi2012-zf} (Fig.~\ref{fig: setup}c).
Two 862-nm IFDLs are used as the fundamentals of the excitation lasers.
As a reference laser, we use an 1,156-nm IFDL~\cite{Takata2019-hj} stabilized to an optical cavity made of an ULE glass with the Pound-Drever-Hall method~\cite{Black2001-bi}.
Then, the repetition rate of the comb is stabilized using the 1,156-nm IFDL, whereas the carrier-envelope offset frequency is stabilized to a 10-MHz reference oscillator using an $f$-2$f$ interferometer method.
Finally, the stability of the comb is transferred to the 862-nm IFDL-1 by offset-locking to the comb, whereas 862-nm IFDL-2 is offset-locked to the 862-nm IFDL-1 device with an offset of an IS between $^{174}\mathrm{Yb}$ and a target isotope.
Note that all the radio-frequency generators relevant to the excitation lasers are referenced to the frequency standard in the National Metrology Institute of Japan via the global positioning system common-view method.

These 862-nm lasers are amplified with TAs, and their wavelengths are converted to 431 nm with second-harmonic generation (SHG).
In this work, we exploit two distinct methods for SHG.
For precision spectroscopy including IS measurements, commercially available wave-guided periodically poled lithium niobate crystals are used.
The typical power at the atomic position is $36$~{\textmu}W and the corresponding peak intensity is 1.8 W cm$^{-2}$.
By contrast, we develop an SHG cavity with a lithium triborate crystal because much more power is desirable for the other experiments, such as the observations of the Rabi oscillation, the relaxation dynamics and the Feshbach resonance.
The typical power and peak intensity at the atomic position are 10 mW and 78 W cm$^{-2}$, respectively.
The 431-nm lasers are delivered to the atom through the same fibre. 
Their powers are stabilized by monitoring the fibre outputs and feeding them back to AOM-1 and AOM-2.
During the excitation, we apply a magnetic field to set the quantization axis and enable the magnetic-field-induced E1 transition.
The strength of the magnetic field is calibrated with the spectroscopy using the $^1S_0 \leftrightarrow \, ^3P_1$ transition~\cite{Budick1967-tt, Baumann1968-hx}.

The number of $\ket{g}$ atoms is measured by the absorption imaging with the $\g \leftrightarrow \: ^1P_1$ transition.
To measure the number of $\ket{e}$ atoms, we first blast the remaining $\ket{g}$ atoms with the same transition as for the absorption imaging, and then shine a 599-nm laser resonant to the $\e \leftrightarrow 4f^{13}5d6s6p \: (J=2)$ transition for repumping the atoms to the $\ket{g}$ state.
We obtain the number of $\ket{e}$ atoms by measuring the number of atoms back in the $\ket{g}$ state after the correction for the finite repumping efficiency with the experimentally measured value of 59\%.

\subsection*{Coherent control of optical clock transition}
\subsubsection*{Measurement procedure}
To determine the M2 transition moment, we utilize the M2 transition mechanism for excitation.
The peak intensity of the excitation laser is $I = 78(7)$ W cm$^{-2}$.
To set the quantization axis, the magnetic field of about 0.19~mT is applied in the $x$-$y$ plane, and the relative angle to the propagation of the excitation laser is $\theta = 36(2)^\circ$.
The polarization of the excitation laser remains along the $z$ axis to satisfy the M2 selection rule.
In addition, we adjust the polarization of all three lattice lasers so that the magic condition holds.
Other parameters are similar to those for the precision spectroscopy.

\subsubsection*{Experimental determination of the M2 transition rate}
When the polarization of the excitation laser is orthogonal to the plane spanned by the propagation and the quantization axis, the matrix element of the $\ket{J_g=0, m_g=0} \leftrightarrow \ket{J_e=2, m_e=0}$ M2 transition is given by~\cite{Trautmann2023-vn}
\begin{eqnarray}
    && \langle J_e, m_e | H_\mathrm{M2} | J_g, m_g \rangle \nonumber \\
    && = \sqrt{\frac{\mu_0 k^2 I}{10 c}} \sin{\theta} \cos{\theta} \langle J_e || Q^{\mathrm{mg}}_{2} || J_g \rangle.
\end{eqnarray}
Here $H_\mathrm{M2}$ denotes the M2 component of the light--matter interaction Hamiltonian in the Coulomb gauge, $\mu_0$ is the vacuum permeability, and $k$ is the wave number of the excitation laser.
$\langle J_e || Q^{\mathrm{mg}}_{2} || J_g \rangle$ is the reduced matrix element of the M2 transition operator.
Using the measured value of $\Omega_{\mathrm{M2}} \equiv \langle J_e, m_e | H_\mathrm{M2} | J_g, m_g \rangle / \hbar = 2 \pi \times 321(3)$~Hz, the M2 transition moment is obtained as $\langle J_e || Q^{\mathrm{mg}}_{2} || J_g \rangle = 1.69(9) \times 10^{-33} \ \mathrm{A} \, \mathrm{m}^3$.
The value in the parenthesis is the $1\sigma$ total uncertainty, dominated by the systematic uncertainty of $I$.
Note that, to obtain an accurate Rabi frequency, we extract the data with excitation time of less than 4 ms in Fig.~\ref{fig: basic}a, and fit them with a sinusoidal function with an exponential damping.

\subsection*{Intrinsic lifetime of the \texorpdfstring{$\e$}{e} state}
\subsubsection*{Measurement procedure}
The temporal evolution of the number of $\ket{g}$ and $\ket{e}$ atoms in a 3D optical lattice is measured as follows.
We prepare $\ket{g}$ atoms in a 3D optical lattice and excite them to the $\ket{e}$ state by applying a $\pi$-pulse with the M2 excitation mechanism.
After the excitation, the remaining $\ket{g}$ atoms are blasted with a resonant $399$-nm laser.
Then, we measure the time dependence of the atom numbers $N_g$ and $N_e$ during a hold time.
In the analysis, the trap loss rate of the $\ket{g}$ state is fixed to the experimentally measured value $\Gamma_g = (70 \ \mbox{s})^{-1}$.

\subsubsection*{Calculation of the BBR quenching rate}
We consider a three-level system: the excited state of the clock transition $\ket{e}$, an upper state $\ket{i} = \ket{J_i, m_i}$ to which $\ket{e}$ is excited by BBR, and a lower state $\ket{q} = \ket{J_q, m_q}$ to which $\ket{i}$ decays.
Assuming that the population in $\ket{i}$ is negligible, the rate equation is given by~\cite{Sutur1997-hu}
\begin{eqnarray}\label{eq: BBR rate}
    \dot{N_q}
    && = \gamma_{e \rightarrow i \rightarrow q} \times N_e  \nonumber \\
    && = \Gamma_{i \rightarrow q} \int \frac{\widetilde{\Omega}_{ei}^2}{4 (\omega - \omega_{ei})^2 + \Gamma_i^2} I(\omega, T) d\omega \times N_e.
\end{eqnarray}
Here $N_{e(q)}$ is the number of $\ket{e(q)}$ atoms.
$\gamma_{e \rightarrow i \rightarrow q}$ is the BBR quenching rate from $\ket{e}$ to $\ket{q}$ via $\ket{i}$.
$\Gamma_{i \rightarrow q}$ is the decay rate from $\ket{i}$ to $\ket{q}$, whereas $\Gamma_i$ is the sum of $\Gamma_{i \rightarrow q}$ over all possible decay channels including the magnetic sublevels of the $\e$ state.
$\omega_{ei}$ is the resonant angular frequency between $\ket{e}$ and $\ket{i}$.
In addition, $\widetilde{\Omega}_{ei}$ is the intensity-normalized Rabi frequency when driven by a monochromatic laser light with the resonance angular frequency $\omega_{ei}$.
The intensity of BBR at temperature $T$ in the frequency range of $[\omega, \omega + d\omega]$ is denoted as $I(\omega, T) d\omega$.

Extended Data Fig.~\ref{fig:lifetime_BBR} shows $\ket{i}$ and $\ket{q}$ states considered in this work.
As for the BBR excitation, we consider four E1 transitions close to the BBR intensity peak at $\SI{10}{\mu}$m: $\ket{e}$ to $^3D_1$, $^3D_2$, $^3D_3$ and $^1D_2$, the transition wavelengths of which correspond to 7.689, 6.396, 4.802 and $\SI{2.228}{\mu}$m, respectively~\cite{Dzuba2018-kc}.
Note that, since these states have the same parity as $\g$, they do not directly decay to the ground state.
As possible $\ket{q}$ states, except magnetic sublevels in the $\e$ state, Ref.~\cite{Porsev1999-ja} considered four low-lying odd-parity states $^3P_0$, $^3P_1$, $^3P_2$ and $^1P_1$.
Summing $\gamma_{e \rightarrow i \rightarrow q}$ for all $\ket{i}$ and $\ket{q}$, the quenching rate from the $\ket{e}$ state is defined as $\gamma_{e} = \sum_{i, q}{\gamma_{e \rightarrow i \rightarrow q}}$.

By measuring the temperature of the heat-generating objects such as coils as well as the temperature of the room with thermistors, the quenching rate is estimated to be $\gamma_{e} = 0.060$~Hz, with contributions of $\gamma_{e \rightarrow \; ^1S_0} = 0.025$~Hz, $\gamma_{e \rightarrow \; ^3P_0} = 0.0028$~Hz and $\gamma_{e \rightarrow \; ^3P_2} = 0.033$~Hz.
Here we define $\gamma_{e \rightarrow \; ^1S_0} \equiv \gamma_{e \rightarrow \; ^3P_1} + \gamma_{e \rightarrow \; ^1P_1}$ since atoms in the $^3P_1$ and $^1P_1$ states decay to the $\g$ state in a much shorter timescale of less than $\SI{1}{\mu}$s and 10~ns, respectively, compared with the observed time constant.
Note that our theoretical estimation shows that the quenching rate to the magnetic sublevels of the $\e$ state is negligibly small.

\subsection*{Interorbital Feshbach resonance}
\subsubsection*{Occupancy-resolved spectroscopy}
To measure the interaction shift $\Delta f$, we perform the occupancy-resolved spectroscopy.
We load ultracold Yb atoms into the 3D isotropic optical lattice at the depth of $45E_r$ per axis, and perform the spectroscopy by applying a $\pi$-pulse of $n=1$ sites.
We adjust the atomic density with evaporation parameters so that there are as many $n=2$ sites but as few $n=3$ sites as possible, since the $n=3$ sites might contaminate the signals of $n=1$ and $n=2$ sites.
Other experimental conditions are similar to those for the coherent control of optical clock transition and the intrinsic lifetime measurement.

\subsubsection*{Determination of the interorbital scattering length \texorpdfstring{$a_{ge}$}{age}}
The frequency shift $\Delta f$ between the $n=1$ and $n=2$ sites is related to the interatomic interaction as~\cite{Kato2013-jz}
\begin{equation} \label{eq: delta f}
    h \Delta f = U_{ge} - U_{gg},
\end{equation}
where $U_{gg(e)}$ represents the interaction energy between $\ket{g}$ and $\ket{g(e)}$.
The interaction energy $U_{gg}$ is given by~\cite{Bloch2008-cq}
\begin{equation} \label{eq: Ugg}
    U_{gg} = \frac{\sqrt{8}}{\pi} k_L a_{gg} E_r s^{3/4}.
\end{equation}
Here $k_L$ is the wave number of the optical lattice, $s$ is the lattice depth in units of $E_r$, and $a_{gg}=5.55812(50)$~nm is the $s$-wave scattering length between $\ket{g}$ atoms~\cite{Borkowski2017-ed}.
Note that the interatomic interaction is well described by the $s$-wave scattering length since the atomic temperature before loading into the optical lattice is as low as about 300 nK.

Combining the measured $\Delta f$ and the calculated $U_{gg}$, $U_{ge}$ is obtained straightforwardly from Eq.~(\ref{eq: delta f}).
To convert it to the $s$-wave scattering length $a_{ge}$, we use the analytic formula for a harmonically trapped atom pair~\cite{Busch1998-xh}:
\begin{equation}\label{eq:a_ge}
    a_{ge} = \frac{a_{\mathrm{ho}}}{\sqrt{2}} \frac{\Gamma(-U_{ge}/(2 \hbar \omega_t) - 1/2)}{\Gamma(-U_{ge}/(2 \hbar \omega_t))},
\end{equation}
where $\omega_t$ is the trap angular frequency, $a_{\mathrm{ho}}$ is the harmonic oscillator length, and $\Gamma(x)$ denotes the gamma function.

\subsection*{IS measurements}
\subsubsection*{Interleaved clock operation}
For each isotope, we measure excitation fractions at the left and right shoulders of atomic spectra, the difference of which is fed back to the AOM frequency as an error signal.
The frequency interval between two shoulders is set constant as $80$~Hz, which is the typical full-width at half-maximum of our spectrum.
We alternately perform this sequence for two isotopes for the interleaved clock operation.
The IS is obtained as the frequency difference between AOM-1 and AOM-2 considering the constant offset frequency between two 862-nm IFDLs.
Note that we estimate and correct a resonance frequency drift of the ULE cavity during the cycle by fitting the time trace of the resulting frequency drift of AOM-1.

\subsubsection*{Evaluation of statistical uncertainties}
To measure the ISs and their statistical uncertainties, the standard condition of the clock operation is set to the experimental condition shown in the ``precision spectroscopy" section.
For each isotope pair, we perform two measurements on different days.
Extended Data Fig.~\ref{fig: IS_time_trace} shows the time traces of IS measurements.
For each measurement, we evaluate the stability of the interleaved clock operation as shown in Fig.~\ref{fig: IS}a, and determine the averaging time until which the overlapping Allan deviation can be fitted well with a line assuming a white frequency noise.
The whole dataset is divided, and the average of each segment is displayed as a blue point (Extended Data Fig.~\ref{fig: IS_time_trace}) after correcting all the systematic effects.
For the error bars, we adopt the overlapping Allan deviation at the averaging time determined above.
The IS is determined as the weighted average of all segments.
For the $1 \sigma$ statistical uncertainty shown as the blue shaded area, we conservatively adopt the larger of the weighted standard error and the error propagation of the error bars.
For comparison, the $1\sigma$ systematic uncertainties are also displayed as orange shaded regions (see Fig.~\ref{fig: IS}c), indicating that the two measurements performed on two distinct days are consistent within the systematic uncertainties for all isotope pairs.

\subsubsection*{Evaluation of systematic effects}
\textit{Lattice light shift.}
The lattice light shift, which is an a.c. Stark shift induced by the optical lattice, gives a major contribution since each isotope has a slightly different magic condition.
Extended Data Fig.~\ref{fig: IS_LLS} summarizes the measured lattice light shifts as a function of the lattice depth for different isotope pairs.
We estimate the lattice light shifts at a lattice depth of $30 E_r$ for the operational condition by fitting the data with the formula considering the nonlinear effect due to the zero-point energy~\cite{Ushijima2018-qc}: 
\begin{equation} \label{eq: LLS}
    f(s) = a \left( s + \frac{\sqrt{s}}{2} \right) + b,
\end{equation}
where $s$ is the lattice depth in units of $E_r$, and $a$ and $b$ are fitting parameters.
Here we suppose that multi-polar effects~\cite{Ushijima2018-qc} are almost common for each isotope.
The uncertainties of the lattice light shifts are evaluated by considering the fitting errors and the lattice depth uncertainty of about 12\%.

\textit{Quadratic Zeeman shift.}
The quadratic Zeeman shift arises from the magnetic-field-induced mixing between the $\ket{e}$ state and others.
Reference~\cite{Dzuba2018-kc} estimated the quadratic Zeeman shift coefficient of the $\gtoe$ transition to be $-4.3 \times 10^5$ Hz T$^{-2}$, corresponding to $-92$~Hz at $14.6$~mT of the current experiment.
If this is common for all isotopes, there would be no systematic shift.
The measurement of the magnetic field dependence of ISs is shown in Extended Data Fig.~\ref{fig: IS_QZS}, indicating no notable dependence on the magnitude of the magnetic field.
This is reasonable since the isotope dependence of the quadratic Zeeman shift is estimated approximately 0.01\% smaller, corresponding to a shift no greater than 10~mHz, much smaller than the statistical uncertainties.
We thus set the correction of the quadratic Zeeman shift as zero and take the $1\sigma$ statistical uncertainties of three or four measurements as the uncertainty.

\textit{Probe light shift.}
The probe light shift is also investigated as shown in Extended Data Fig.~\ref{fig: IS_PLS}, indicating no systematic dependence of ISs on the excitation laser intensities.
Thus, we evaluate the corrections and uncertainties in a similar way to that for the quadratic Zeeman shift.

\textit{Servo error.}
The servo error in the clock operation is evaluated with the mean values and the overlapping Allan deviations of the excitation fractions on both shoulders of the excitation profile.

\textit{BBR shift.}
In Ref.~\cite{Dzuba2018-kc}, the BBR shift at the temperature of $300$~K is estimated as $0.2$~Hz, which is, in principle, a common perturbation for different isotopes, similar to the quadratic Zeeman shift and the probe light shift.
However, the evaporation time heavily depends on isotopes, causing the possible change of the temperature of coils surrounding the atoms.
We measure the temperatures of the coils with thermistors and evaluate the isotope dependence of the BBR shift with the calculated BBR coefficient~\cite{Dzuba2018-kc}.
Consequently, the BBR shift turns out to be well below $1$~mHz.

\textit{AOM chirp.}
The AOM chirp is caused by a phase shift that arises when the excitation laser is switched on and off.
Since we employ the same products for AOM-1 and AOM-2 as Ref.~\cite{Ono2022-oy}, we adopt the same value $9(6)$~mHz.

\subsection*{Atomic calculations for electronic factors}
The relativistic calculations for the electronic factors ($F_i,$ \Gtwo and $X_i$) are carried out using the DIRAC code~\cite{saue2020dirac,DIRAC22}. The equation-of-motion coupled cluster (EOM-CC) method~\cite{shee2018equation} is used for the transitions $\g \leftrightarrow \:$\Dtw, $^2S_{1/2} \leftrightarrow \:$\DD, $^2S_{1/2} \leftrightarrow \:$\Dth and $\g \leftrightarrow \:$\PP\ transitions, and generalized-active-space configuration interaction (GASCI) method~\cite{Fleig2003JCP,Knecht2010JCP} is used for the $\g\: \leftrightarrow \:$\JJ\ and $^2S_{1/2} \leftrightarrow \:$\FF\ transitions.
$F_i$ and $G^{(2)}_i$ are calculated using the existing property module in the DIRAC code. The one-electron integral for the Yukawa potential is newly implemented in this study. The uncertainties of the calculations are estimated from the effect of various computational parameters. The details are provided in Supplementary Information, section \RomanNumeralCaps{4}. The calculated electronic factors are summarized in Extended Data Table~\ref{tab: electronic factor} and Extended Data Fig.~\ref{fig: X}.

\subsection*{King plot analysis}
\subsubsection*{Constraint on the new particle by the 3D generalized King plot}
To obtain the favoured region of $|y_e y_n|$ from the 3D generalized King plot, shown by the blue solid line and shaded area in Fig.~\ref{fig: KP}c, we consider the two leading-order SM terms, one higher-order SM term and the PS.
The relation between ISs of three transitions is then given by
\begin{eqnarray} \label{eq: alpha_NP}
    \dn{3}{AA_0} = k \w{AA_0} + f_1 \dn{1}{AA_0} + f_2\dn{2}{AA_0} \nonumber\\
    + \alpha_{\mathrm{NP}} (X_3 - f_1 X_1 - f_2 X_2) (A - A_0),
\end{eqnarray}
where $k$, $f_1$, $f_2$ and $\alpha_{\mathrm{NP}}$ are fitting parameters.
By a fit to the data of $\w{AA_0}$ and $\dn{i}{AA_0} \ (i=1,2,3)$, the central value and the $95\%$ confidence interval of $\alpha_{\mathrm{NP}}$ is obtained.

\subsubsection*{QFS-subtracted IS}
Assuming that an IS consists of the two leading-order SM terms and the two higher-order field shift terms, ISs of four transitions and four isotope pairs can be written in the matrix form as $V = EN$, where
\begin{equation}
    V = 
    \begin{pmatrix}
    \dn{1}{A_1A_0} & \dn{1}{A_2A_0} & \dn{1}{A_3A_0} & \dn{1}{A_4A_0} \\
    \dn{2}{A_1A_0} & \dn{2}{A_2A_0} & \dn{2}{A_3A_0} & \dn{2}{A_4A_0} \\
    \dn{3}{A_1A_0} & \dn{3}{A_2A_0} & \dn{3}{A_3A_0} & \dn{3}{A_4A_0} \\
    \dn{4}{A_1A_0} & \dn{4}{A_2A_0} & \dn{4}{A_3A_0} & \dn{4}{A_4A_0} \\
\end{pmatrix}, 
\end{equation}
\begin{equation}
    E = 
    \begin{pmatrix}
    K_1 & F_1 & G^{(4)}_1 & G^{(2)}_1 \\
    K_2 & F_2 & G^{(4)}_2 & G^{(2)}_2 \\
    K_3 & F_3 & G^{(4)}_3 & G^{(2)}_3 \\
    K_4 & F_4 & G^{(4)}_4 & G^{(2)}_4 \\
\end{pmatrix}, 
\end{equation}
\begin{eqnarray}
    && N =  \\
    && \begin{pmatrix}
    \w{A_1A_0} & \w{A_2A_0} & \w{A_3A_0} & \w{A_4A_0} \\
    \dr{A_1A_0} & \dr{A_2A_0} & \dr{A_3A_0} & \dr{A_4A_0} \\
    \drr{A_1A_0} & \drr{A_2A_0} & \drr{A_3A_0} & \drr{A_4A_0} \\
    \ddr{A_1A_0} & \ddr{A_2A_0} & \ddr{A_3A_0} & \ddr{A_4A_0} \\
    \end{pmatrix}. \nonumber
\end{eqnarray}
To subtract the QFS, we define a matrix $W$ with theoretically calculated $G_i^{(2)}$ ratios:
\begin{equation}
    W = 
    \begin{pmatrix}
    1 & 0 & 0 & 0 \\
    -G^{(2)}_2 / G^{(2)}_1 & 1 & 0 & 0 \\
    -G^{(2)}_3 / G^{(2)}_1 & 0 & 1 & 0 \\
    -G^{(2)}_4 / G^{(2)}_1 & 0 & 0 & 1 \\
    \end{pmatrix}.
\end{equation}
By multiplying $W$ from the left, the formula is transformed as $\widetilde{V} = \widetilde{E} N$.
Here $\widetilde{V}$ consists of the QFS-subtracted IS $\widetilde{\dn{}{}}$ defined in Eq.~(\ref{eq: QFS-subtracted IS}):
\begin{equation}
    \widetilde{V} = 
    \begin{pmatrix}
    \dn{1}{A_1A_0} & \dn{1}{A_2A_0} & \dn{1}{A_3A_0} & \dn{1}{A_4A_0} \\
    \sdn{2}{A_1A_0} & \sdn{2}{A_2A_0} & \sdn{2}{A_3A_0} & \sdn{2}{A_4A_0} \\
    \sdn{3}{A_1A_0} & \sdn{3}{A_2A_0} & \sdn{3}{A_3A_0} & \sdn{3}{A_4A_0} \\
    \sdn{4}{A_1A_0} & \sdn{4}{A_2A_0} & \sdn{4}{A_3A_0} & \sdn{4}{A_4A_0} \\
\end{pmatrix}.
\end{equation}
$\widetilde{E}$ is given by
\begin{equation}
    \widetilde{E}
    = \begin{pmatrix}
    \widetilde{K_1} & \widetilde{F_1} & \widetilde{G}^{(4)}_1 & G^{(2)}_1 \\
    \widetilde{K_2} & \widetilde{F_2} & \widetilde{G}^{(4)}_2 & 0 \\
    \widetilde{K_3} & \widetilde{F_3} & \widetilde{G}^{(4)}_3 & 0 \\
    \widetilde{K_4} & \widetilde{F_4} & \widetilde{G}^{(4)}_4 & 0 \\
    \end{pmatrix}, 
\end{equation}
clearly indicating that the QFS is subtracted.
Generally, by utilizing one reference transition and the theoretical values of electronic factor ratios, one higher-order source can be eliminated, and the effective degree of freedom increases by one.

To obtain the information of the new particle with the QFS-subtracted ISs $\sdn{i}{A'A}$, $X_i$ should also be modified as
\begin{equation}
    \widetilde{X_i} = X_i - \frac{G^{(2)}_i}{G^{(2)}_{i_0}} X_{i_0}.
\end{equation}
Replacing $\dn{}{}$ and $X_i$ in Eq.~(\ref{eq: alpha_NP}) with their QFS-subtracted counterparts, the product of the coupling constants is obtained in a similar way as in the previous section.

\subsubsection*{\texorpdfstring{$G^{(2)}_i$}{G2} ratios to minimize the nonlinearity}
We determine $G^{(2)}_i$ ratios to minimize $\chi^2$ in the 3D generalized King plot with the QFS-subtracted ISs $\sdn{}{}$ as follows.
By defining the inverse of $E$ as 
\begin{equation} \label{eq: alpha of original KP}
\begin{split}
    E^{-1} 
    \equiv
    \begin{pmatrix}
        \beta_1 & \beta_2 & \beta_3 & \beta_4 \\
        \hdotsfor{4} \\
        \hdotsfor{4} \\
        \hdotsfor{4} \\
    \end{pmatrix}
\end{split},
\end{equation}
we obtain $N = E^{-1} V$, where the first row corresponds to the 4D generalized King linearity and allow us to determine $\beta_i \ (i=1,2,3,4)$ from the fit.
From $E^{-1}E=I_4$, where $I_4$ is the $4\times4$ identity matrix, the $G^{(2)}_i$ ratios should satisfy the following equation:
\begin{equation} \label{eq: G2 ratio for zero chi2}
    \frac{G^{(2)}_4}{G^{(2)}_1} = \beta_1'  + \beta_2' \frac{G^{(2)}_2}{G^{(2)}_1} + \beta_3' \frac{G^{(2)}_3}{G^{(2)}_1},
\end{equation}
where $\beta_i' = -\beta_i/\beta_4 \: (i=1,2,3)$.
$G^{(2)}_i / {G^{(2)}_{578}} \ (i=431, 411, 467)$ are determined as a point such that it is on the 3D plane by Eq.~(\ref{eq: G2 ratio for zero chi2}) and has the shortest distance from the calculated value in Extended Data Table~\ref{tab: electronic factor}, yielding $G^{(2)}_i / {G^{(2)}_{578}} = -1.804, 1.584,$ and $-3.676$ for $i=431, 411,$ and 467, respectively.
These values are within the theoretical uncertainties, suggesting that the QFS is one of the sources contributing to the measured ISs within the current measurement precision.

\subsubsection*{Projection of 4D King plot}
We present how to set the prospected bound by the 4D King plot approach with an unstable isotope of $^{166}$Yb.
Since the atomic masses of $^{172}$Yb and $^{166}$Yb were measured~\cite{Wang2021-ar}, the central value of $\w{166, 176}$ is easily obtained by converting atomic masses to nuclear masses $m_A$ (see Appendix A in Ref.~\cite{Ono2022-oy}).
We assume its uncertainty in the future measurement could be improved at the same level as for stable isotope pairs, namely $4 \times 10^{-12}$~\cite{Door2025-jv}.
$\dn{i}{166, 176} \ (i=578, 431, 411)$ are estimated using the two-dimensional King relation between the $^1S_0 \; \leftrightarrow \: ^3P_1$ transition~\cite{Fricke2004}.
Then, $\dn{467}{166, 176}$ is obtained to satisfy the 4D King relation, which holds for four stable isotope pairs.
The uncertainties of four $\dn{}{166, 176}$ are set to be 10 Hz, similar to those of stable isotope pairs.
In addition, we suppose that one certain higher-order SM term mostly causes the nonlinearity observed in the 3D King plot and others do not contribute within the measured uncertainties, meaning that the 4D King linearity holds.
Consequently, the projected upper bound is set in a similar way as for the 3D generalized King plot, as shown by the red dotted line in Fig.~\ref{fig: KP}c.

\subsection*{Dual King plot analysis}
\subsubsection*{Generalization of the dual King linearity}
Here we briefly summarize the basic idea of the dual King plot~\cite{Yamamoto2023-zq} and derive Eqs.~(\ref{subeq: dual IS})-(\ref{subeq: dual ddr}).
Considering four isotope pairs, four transitions, and the four SM terms, ISs are written as $V^T = N^T E^T$.
Here let us define the inverse of $N^T$ as
\begin{equation}
    \left( N^T \right)^{-1} =
    \begin{pmatrix}
        \hdotsfor{4} \\
        \hdotsfor{4} \\
        \hdotsfor{4} \\
        \alpha_1 & \alpha_2 & \alpha_3 & \alpha_4 \\
    \end{pmatrix},
    \label{eq: dual 4D 2}
\end{equation}
and obtain $\left(N^T\right)^{-1} V^T = E^T$.
By focusing on the linear relation to the $G^{(2)}_i$ column in $E^T$ on the right-hand side, the linear relationship between ISs is obtained as Eq.~(\ref{subeq: dual IS}).
Eqs.~(\ref{subeq: dual mu})-(\ref{subeq: dual ddr}) are obtained from the forth row of the matrix relation $\left(N^T\right)^{-1} N^T = I_4$.

\subsubsection*{Constraints on nuclear factors}
In setting the constraints on $\dr{}$ and $\drr{}$, to account for the uncertainties of the atomic calculation for $G^{(2)}_i$, we employ the following procedure.
First, the value of $G^{(2)}_i$ is sampled using the Monte Carlo method, where a uniform distribution with the width of the uncertainty, shown in Extended Data Table ~\ref{tab: electronic factor}, is considered. 
Then, the coefficients $\alpha_m \ (m=1,2,3,4)$ of Eqs.~(\ref{subeq: dual dr}) and (\ref{subeq: dual drr}) are obtained using Eq.~(\ref{subeq: dual IS}) and (\ref{subeq: dual mu}).
We iterate the above procedure 1000 times to construct the histogram for $\alpha_m$.
Then, the best-fitted plane is obtained as the highest probability plane, while the $68 \%$ confidence region is used for the uncertainty of the plane.

The distance from the plane shown in Fig.~\ref{fig: dual KP} is given as follows. 
For convenience, let us consider a point located at coordinates $(x, y, z)$ in the nuclear-factor-ratio space.
Here $z$ axis corresponds to $\eta^{168, 176}/\eta^{174, 176}$, where $\eta \in \left( \dr{}, \drr{} \right)$. 
The normalized distance is defined as the residual in units of the $1 \sigma$ uncertainty along the $z$ axis. 
The error bar is defined as the interval to cover the region associated with the $1\sigma$ interval of the point. 

\renewcommand{\figurename}{Extended Data Fig.}
\renewcommand{\tablename}{Extended Data Table}
\setcounter{figure}{0}
\setcounter{table}{0}
\onecolumngrid
\clearpage

\begin{figure*}[!t]
\centering
\includegraphics[width = 0.9\linewidth]{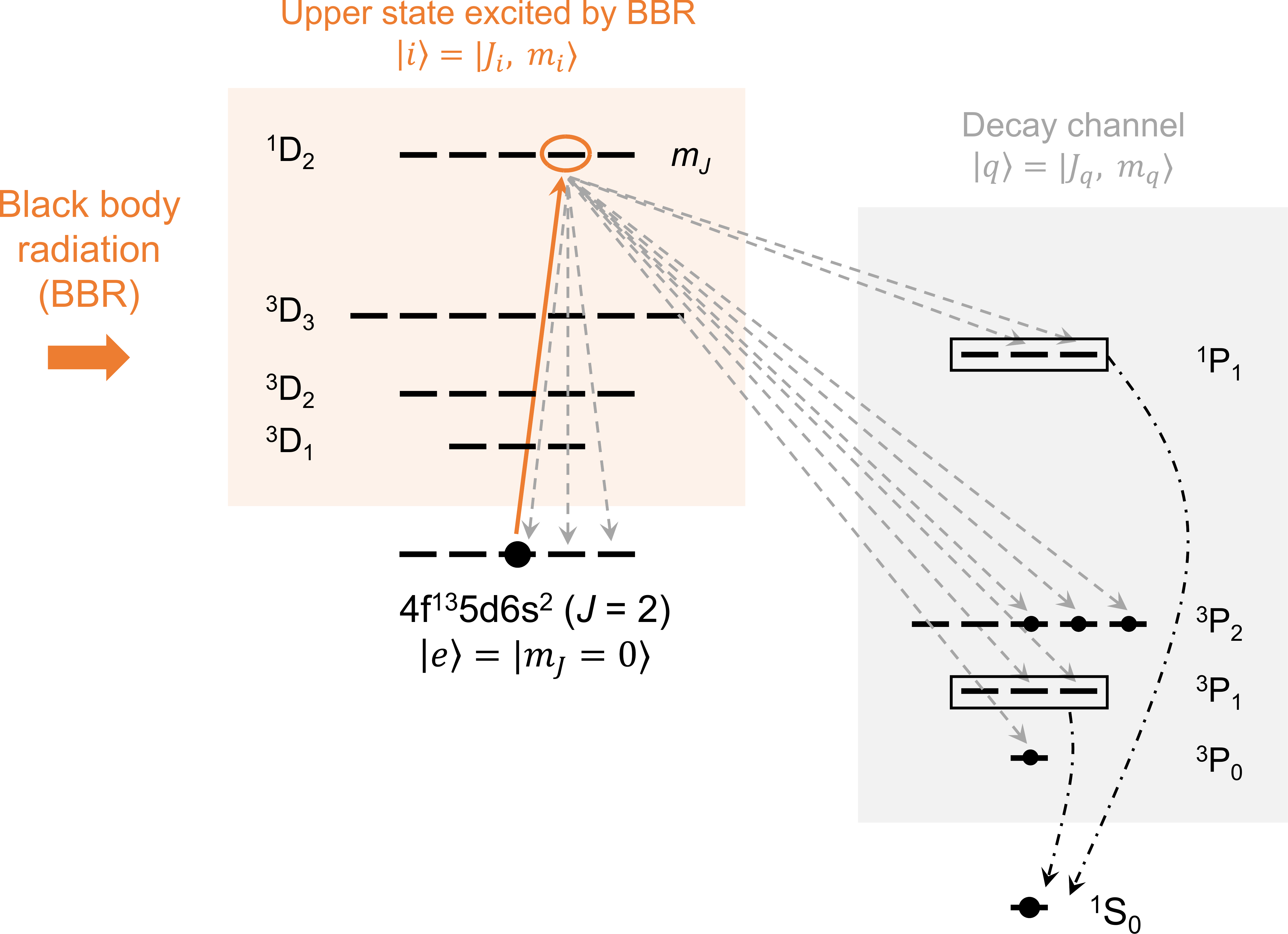}
\caption{Schematic diagram of the BBR quenching of the $\ket{e}$ state. In this figure, we show the case where the $\ket{e}$ state is excited to $\ket{i} = \: ^1D_2 \ (m_J = +1)$ by BBR.}
\label{fig:lifetime_BBR}
\end{figure*}
\clearpage

\begin{figure*}[!h]
\centering
\includegraphics[width = 0.7\linewidth]{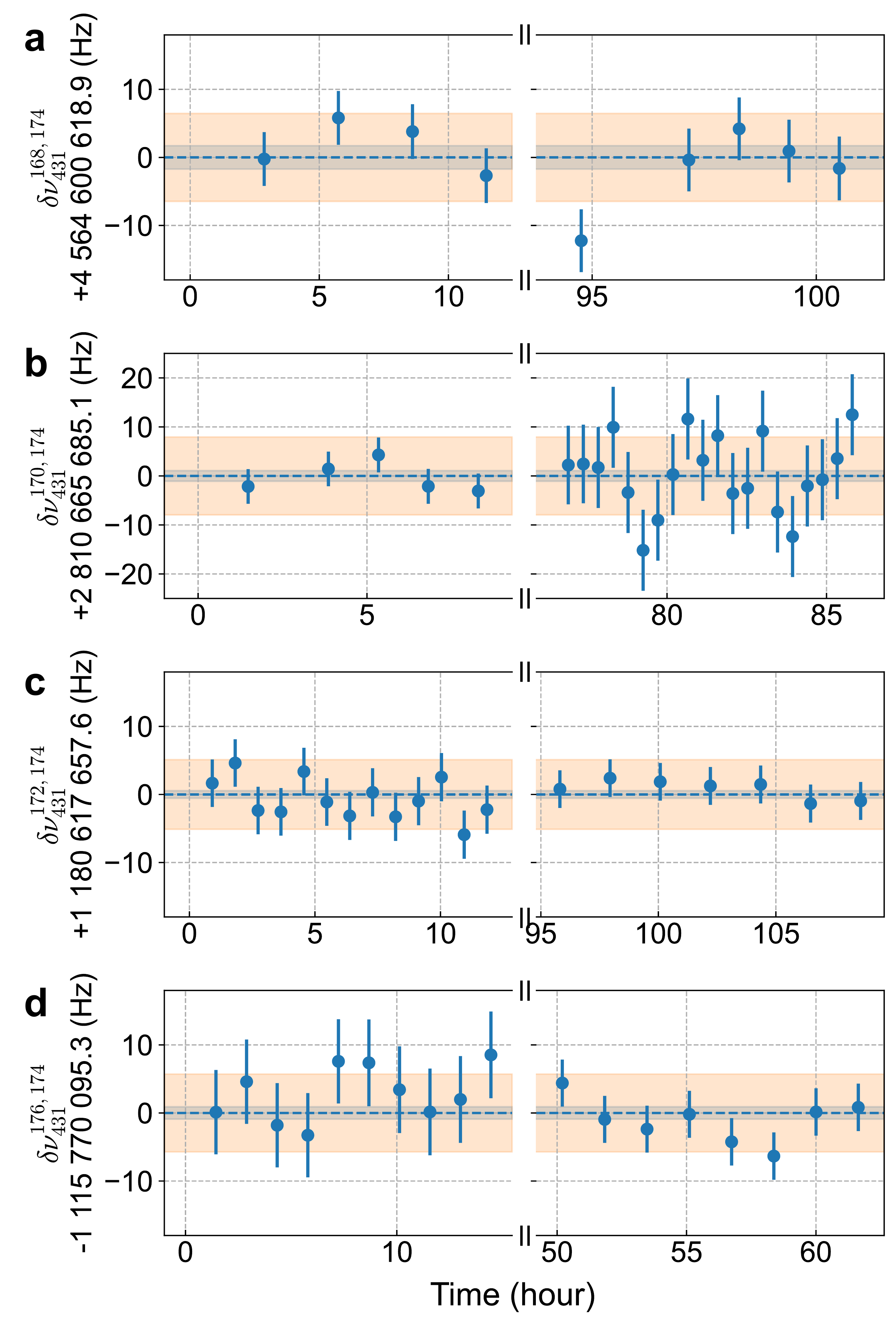}
\caption{Time traces of IS measurements. Each panel presents the result for a specific isotope pair: $^{168}$Yb--$^{174}$Yb (\textbf{a}), $^{170}$Yb--$^{174}$Yb (\textbf{b}), $^{172}$Yb--$^{174}$Yb (\textbf{c}) and $^{176}$Yb--$^{174}$Yb (\textbf{d}). A blue data point is the mean value of each segment after the systematic correction, while an error bar shows the overlapping Allan deviation. Blue (orange) shaded regions represent the $1\sigma$ statistical (systematic) uncertainties. Note that \textbf{d} is the same as Fig.~\ref{fig: IS}b.}
\label{fig: IS_time_trace}
\end{figure*}
\clearpage

\begin{figure*}[!h]
\centering
\includegraphics[width = 0.7\linewidth]{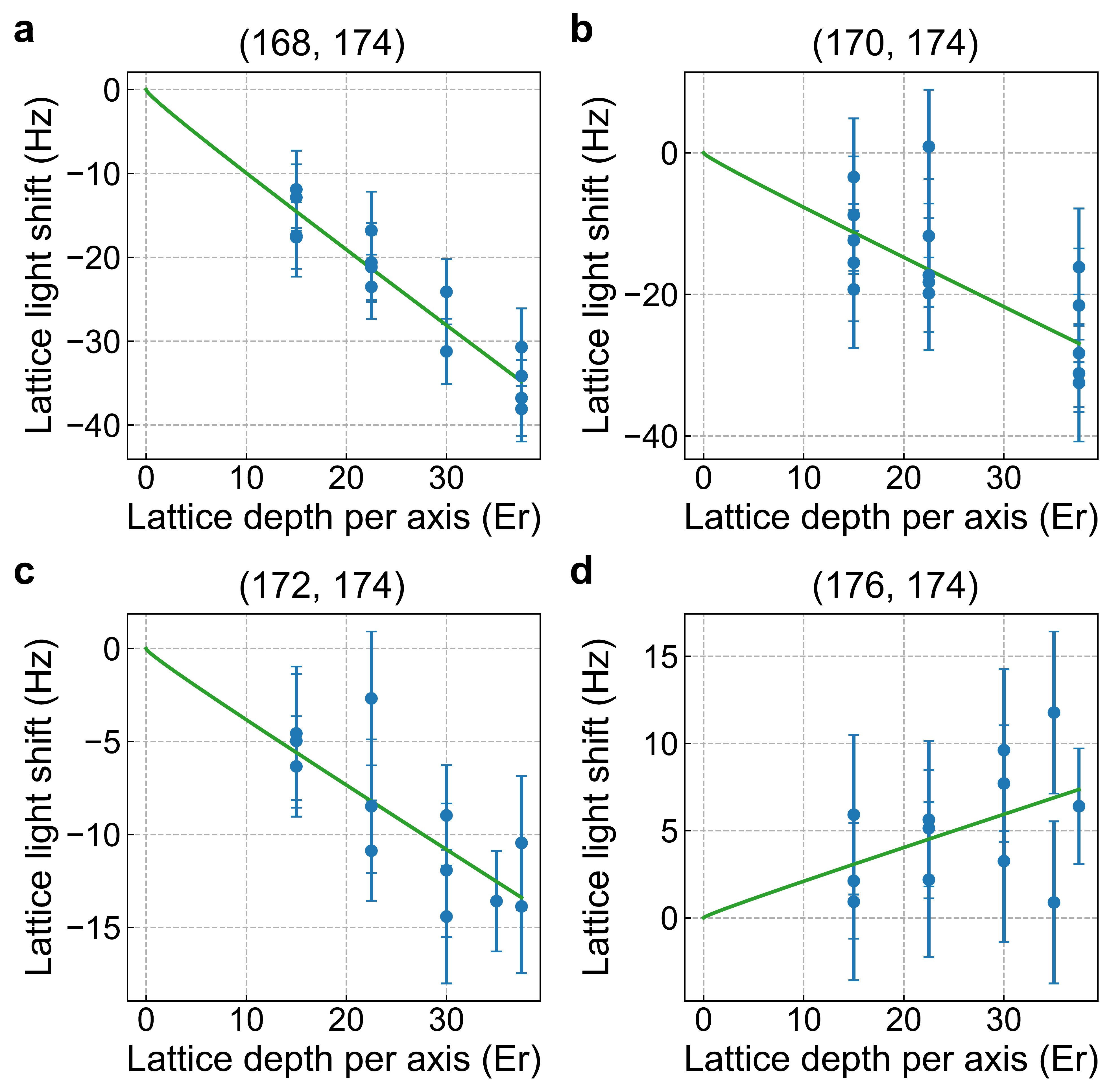}
\caption{Lattice light shift. The measured ISs $\dn{431}{A, 174}$ are plotted with respect to the lattice depth per axis, where $A =$ 168 (\textbf{a}), 170 (\textbf{b}), 172 (\textbf{c}) and 176 (\textbf{d}). Points and error bars denote the mean values and the $1\sigma$ statistical uncertainties, respectively, with the latter derived from the overlapping Allan deviations. Green solid lines are fitting curves by Eq.~(\ref{eq: LLS}) and the determined $y$ axis intercepts are subtracted from the data. During the measurement, we keep the lattice depth ratio for the three axes. To satisfy the Lamb-Dicke condition, the lattice depth per axis is set to be more than $15 E_r$, at which the Lamb-Dicke factor is $0.66$.}
\label{fig: IS_LLS}
\end{figure*}
\clearpage

\begin{figure*}[!h]
\centering
\includegraphics[width = 0.7\linewidth]{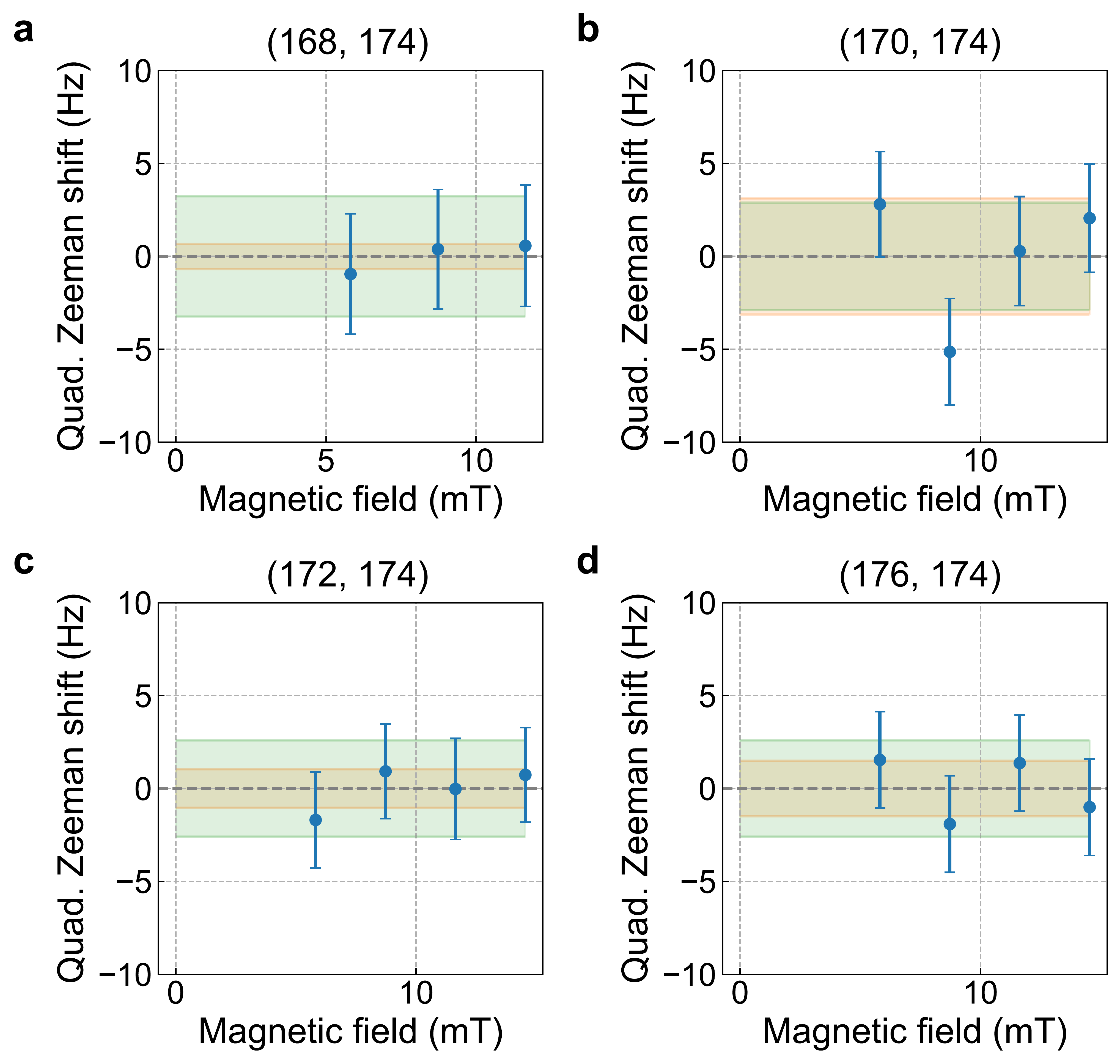}
\caption{Investigation of the quadratic Zeeman shift. The magnetic field dependences of $\dn{431}{A, 174}$ are shown, where $A =$ 168 (\textbf{a}), 170 (\textbf{b}), 172 (\textbf{c}) and 176 (\textbf{d}). Points and error bars denote the mean values and the $1\sigma$ statistical uncertainties, respectively, with the latter derived from the overlapping Allan deviations. Green shaded regions represent the average of error bars, while orange shaded areas denote the standard deviation of three or four data points.}
\label{fig: IS_QZS}
\end{figure*}
\clearpage

\begin{figure*}[!h]
\centering
\includegraphics[width = 0.7\linewidth]{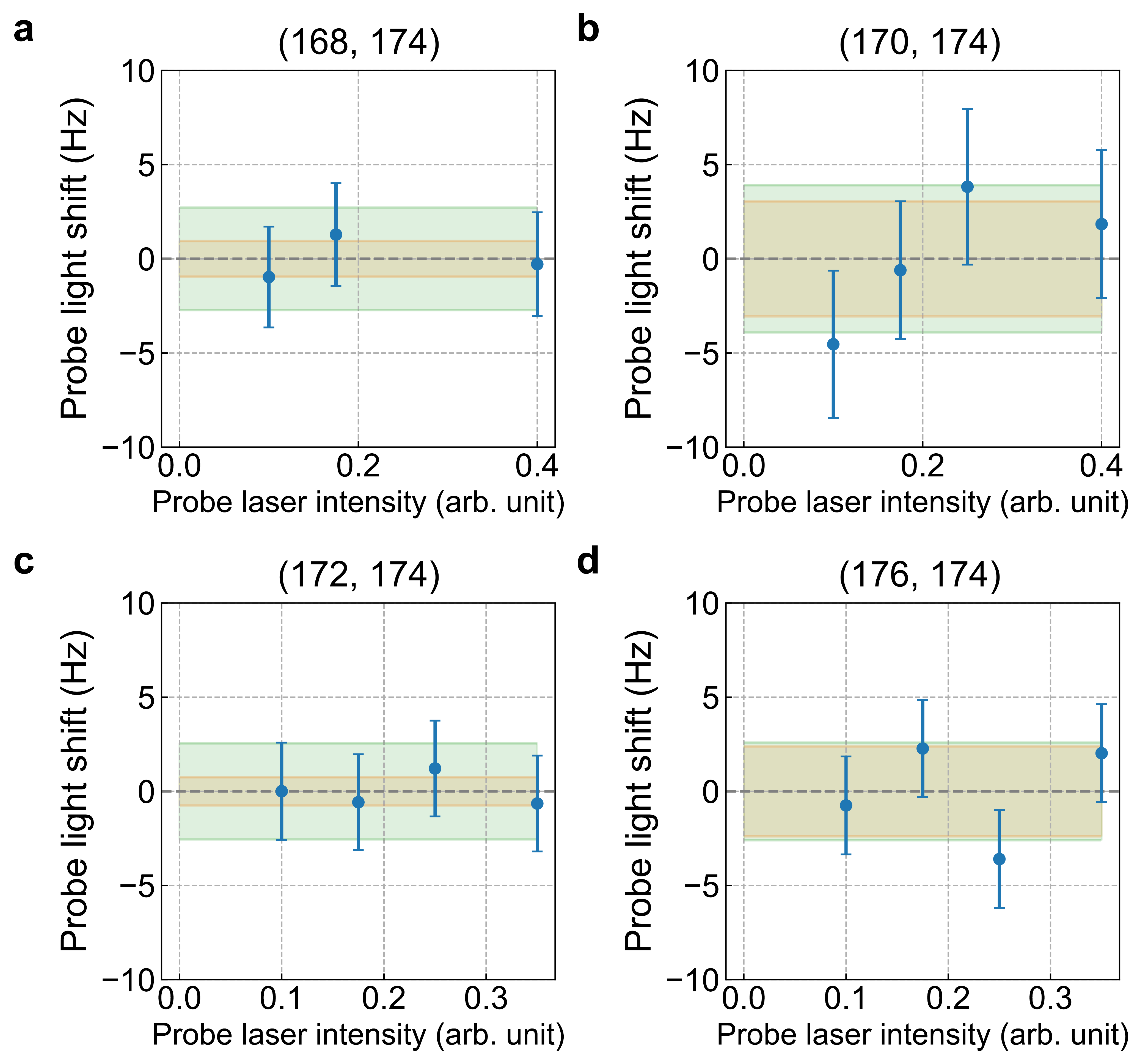}
\caption{Investigation of the probe light shift. The probe laser intensity  dependences of $\dn{431}{A, 174}$ are shown, where $A =$ 168 (\textbf{a}), 170 (\textbf{b}), 172 (\textbf{c}) and 176 (\textbf{d}). Points and error bars denote the mean values and the $1\sigma$ statistical uncertainties, respectively, with the latter derived from the overlapping Allan deviations. The definitions of the two shaded areas are the same as Extended Data Fig.~\ref{fig: IS_QZS}.}
\label{fig: IS_PLS}
\end{figure*}
\clearpage

\begin{figure*}[!t]
\centering
\includegraphics[width = 0.7\linewidth]{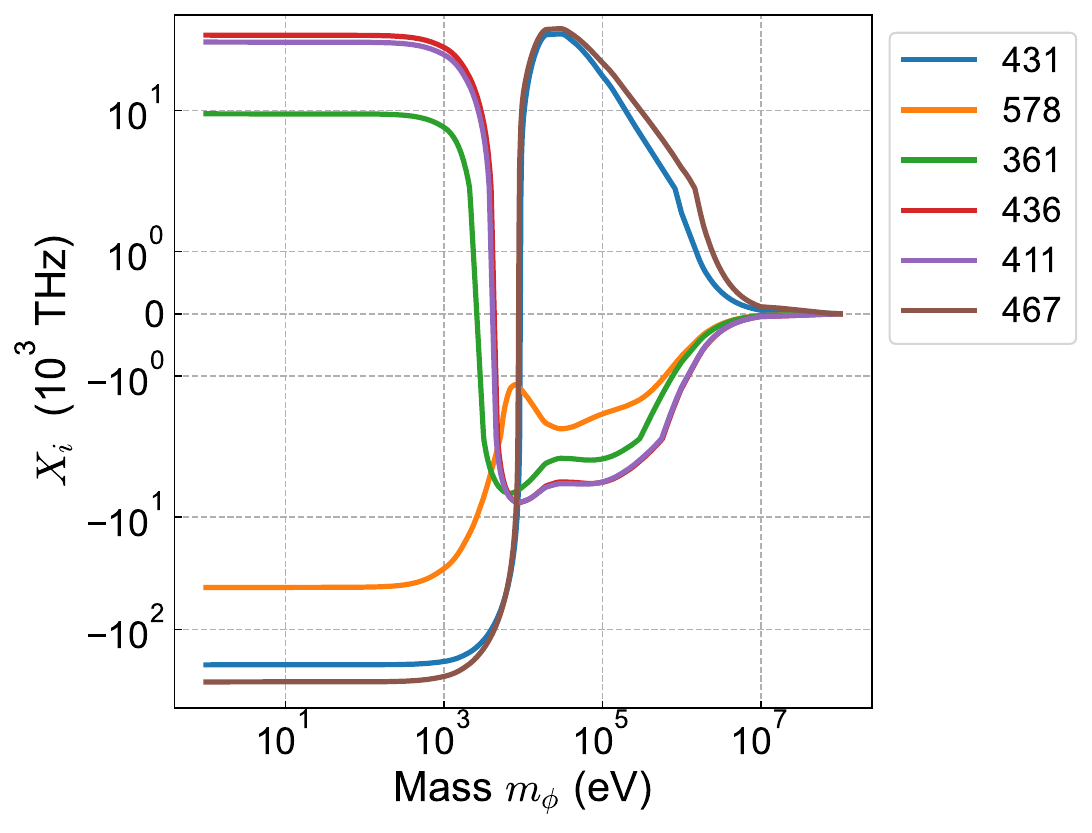}
\caption{PS electronic factor $X_i$ for six transitions. The calculated values of $X_i$ are plotted as a function of the new-particle mass $m_{\phi}$.}
\label{fig: X}
\end{figure*}
\clearpage

\begin{table*}[!t]
\tabcolsep = 0.2cm
    \centering
    \begin{tabular}{c SSS}
    \hline
        Transition & {$F_i$ (GHz/fm$^2$)} & {$G^{(2)}_i$ (MHz/fm$^4$)} & {$X_i$ ($\times 10^3$ THz)} \\
        \hline
        361 &  &  & 9.3176 \\
        431 & 19.3(0.5) & -54.5(1.4) & -206.53 \\
        578 & -10.4(0.2) & 29.9(0.5) & -42.333 \\
        411 & -17.0(0.3) & 48.6(1.2) & 40.545 \\
        436 &  &  & 46.749 \\
        467 & 38.4(1.0) & -109(3) & -293.02 \\
    \hline
    \end{tabular}
    \caption{Summary of the calculated $F_i$, $G^{(2)}_i$, and $X_i$ at $m_{\phi}=1$~eV. The estimated $1\sigma$ uncertainties of the calculation are provided in parentheses. `361' and `436' refer to the transitions  $\g\: \leftrightarrow \:$ \Dtw\ and $^2S_{1/2} \: \leftrightarrow \:$ \Dth, respectively. The mass dependence of $X_i$ is displayed in Extended Data Fig.~\ref{fig: X}.}
    \label{tab: electronic factor}
\end{table*}
\clearpage

\end{document}


\title{Orders-of-magnitude improved precision spectroscopy\\of an inner-shell orbital clock transition in neutral ytterbium\\ - Supplementary Information}

\maketitle

\thispagestyle{empty}
\onecolumngrid
\tableofcontents
\clearpage

\twocolumngrid

\section{Precise determination of the magic condition}
\begin{figure}[!tb]
\centering
\includegraphics[width=0.95\linewidth]{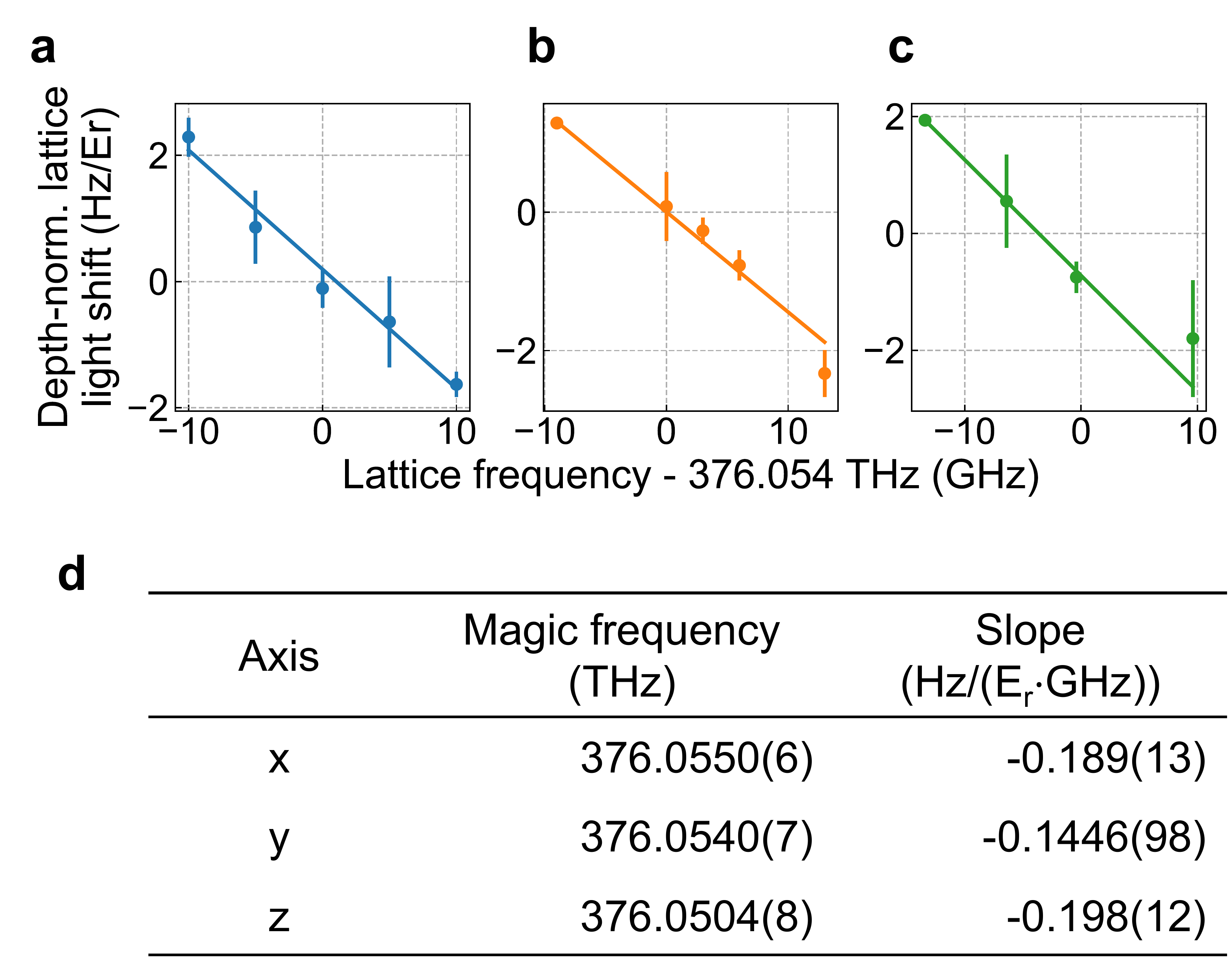}
\caption{Determination of the magic condition. \textbf{a-c}, Lattice frequency dependence of the depth-normalized lattice light shift. \textbf{a}, \textbf{b}, and \textbf{c} correspond to the $x$, $y$, and $z$-axis, respectively. The data points and error bars represent the mean values and $1\sigma$ statistical uncertainties, respectively, derived from the linear fit. The solid lines are the linear fit to the data. \textbf{d}, Magic frequencies and slopes of the depth-normalized differential lattice light shifts. The values in parentheses represent the $1\sigma$ fitting uncertainties.}
\label{fig: magic}
\end{figure}

The magic condition is crucially important in the precision measurement using optically trapped atoms~\cite{Katori2003-hb}.
Two magic wavelengths of $797.2(4)$ and $834.2(1)$ nm were reported for the condition that the lattice laser polarization is perpendicular to the quantization axis~\cite{Ishiyama2023-tv}.
To accurately determine a possibly-axis-dependent magic frequency, we perform interleaved spectroscopy of $^{174}$Yb with three lattice depths on each axis, while the lattice depths of the other two axes remain constant.
We apply a magnetic field of 14.6~mT along the $z$-axis, and the polarizations of the lattice lasers are set perpendicular to the quantization axis, as shown in Fig. 1b in the main text.
The excitation laser is applied for 200 ms to perform spectroscopy.
By repeating measurements with different frequencies of the titanium-sapphire laser, the lattice light shift $\Delta \nu$ is obtained as a function of the lattice depth difference.
From a linear fit of the obtained $\Delta \nu$ with the lattice depth, we obtain the lattice light shift normalized with the recoil energy $E_\mathrm{r}$ as a function of the lattice frequency, as shown in Figs.~\ref{fig: magic}a-c.
From a linear fit to the data, we obtain the magic frequency and the slope for each axis as summarized in Fig.~\ref{fig: magic}d.
Note that the magic condition of the $z$-axis deviates from those of the other two axes, which might be explained by a tensor light shift that depends on a relative angle between the lattice laser's polarization vector and the quantization axis.
In the precision measurement in this work, the lattice laser frequency is set to the magic frequency of the $y$-axis, which is parallel to the excitation laser.
The corresponding wavelength is 797.206(2) nm, indicating an improvement of about two orders of magnitude compared to the previous work~\cite{Ishiyama2023-tv}.

\section{Experimental determination of the magnetic field-induced E1 transition strength}
\begin{figure*}[!tb]
\centering
\includegraphics[width=0.8\linewidth]{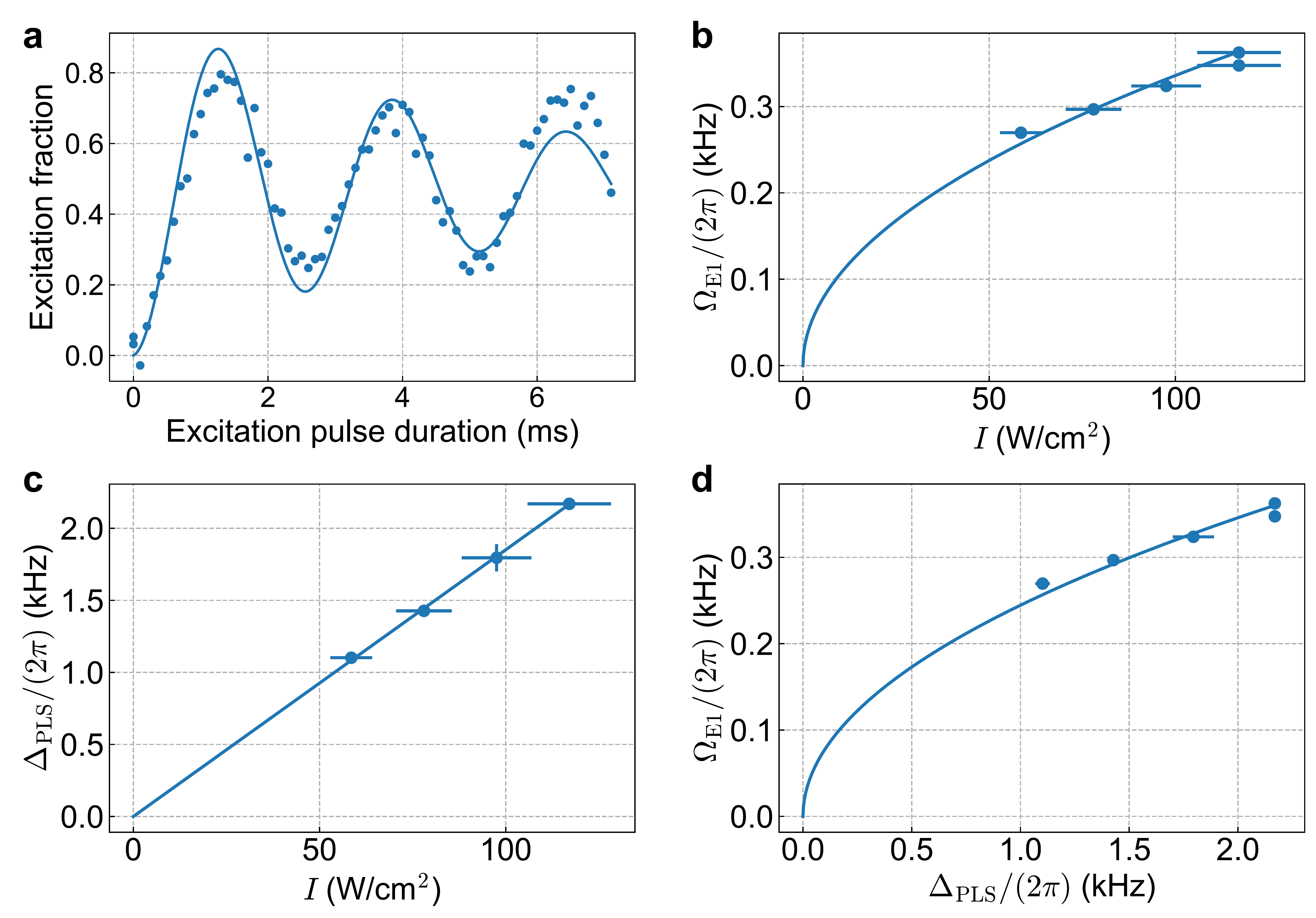}
\caption{Experimental determination of the magnetic field-induced E1 transition strength. 
\textbf{a}, Coherent Rabi oscillation. The solid line is a fitting curve by a sinusoidal function with an exponential damping. The Rabi frequency and the $1/e$ decay time are obtained as $\Omega_\mathrm{E1} = 2 \pi \times 387(2)$~Hz and $5.5(5)$~ms, respectively, where the values in the parentheses are the $1\sigma$ fitting uncertainties. 
\textbf{b}, Rabi frequency $\Omega_\mathrm{E1}$ as a function of $I$. The data points indicate the mean values derived from the fit, while the vertical error bars associated with the $1\sigma$ fitting uncertainties are smaller than the symbol sizes. The horizontal error bars represent the systematic uncertainties of $I$. The solid line is a fitting curve with a function of a square root scaling. Here, to obtain an accurate Rabi frequency, only the first cycle of the Rabi oscillation is extracted and fitted with a sinusoidal function with an exponential damping (e.g., up to 2.9~ms in \textbf{a}).
\textbf{c}, Probe light shift $\Delta_\mathrm{PLS}$ with respect to the intensity $I$. The data points and vertical error bars are the mean values and $1\sigma$ statistical uncertainties, respectively, derived from the fit. The solid line is a linear fit to the data.
\textbf{d}, Investigation of the relative transition strength $\xi \equiv \Omega_{\mathrm{E1}}^2/\Delta_{\mathrm{PLS}}$. The same function as for \textbf{b} is used to fit the data.}
\label{fig: E1}
\end{figure*}

Here, we report the observation of the coherent Rabi oscillation and the determination of the corresponding transition strength for the magnetic field-induced E1 transition.
Figure~\ref{fig: E1}a displays the Rabi oscillation with the excitation laser intensity of $I=117(11)$~W/cm$^2$ at a magnetic field of $B=14.64(14)$~mT.
We investigate the intensity dependence of the Rabi frequency $\Omega_{\mathrm{E1}}$, as shown in Fig.~\ref{fig: E1}b.
By fitting the data with a function of a square root scaling, $\Omega_{\mathrm{E1}}$ normalized with $I$ is obtained as $\Omega_{\mathrm{E1}} / \sqrt{I} = 2\pi \times 1.06(5)$~Hz/(mW/cm$^2$)$^{1/2}$, corresponding to $\Omega_{\mathrm{E1}} / (\sqrt{I} B) = 2\pi \times 73(4)$~Hz/T(mW/cm$^2$)$^{1/2}$.
The values in the parentheses are the total uncertainties, dominated by the systematic uncertainty of $I$.

We also investigate the relative excitation strength $\xi \equiv \Omega_{\mathrm{E1}}^2 / \Delta_{\mathrm{PLS}}$, where $\Delta_{\mathrm{PLS}}$ denotes the probe light shift ~\cite{Lange2021-tx, Klusener2024-yh}.
Figure~\ref{fig: E1}c shows the intensity dependence of $\Delta_\mathrm{PLS}$.
From a linear fit, we obtain $\Delta_{\mathrm{PLS}} = 2\pi \times 0.0185(19)$~Hz/(mW/cm$^2$)$^{1/2}$, whose accuracy is also limited by the $I$ uncertainty.
Here, we perform the spectroscopy with the intensity of $I=117(11)$~W/cm$^2$ at the beginning and the end of the measurement, and the observed resonance shift is used to correct the resonance frequency drift of the ULE cavity during the measurement.
Figure~\ref{fig: E1}d illustrates the relation between $\Delta_{\mathrm{PLS}}$ and $\Omega_{\mathrm{E1}}$.
By fitting the data with the same function as for Fig.~\ref{fig: E1}b, $\xi$ is determined as $2\pi \times 59.8(1.8)$~Hz, where the value in the parenthesis is a fitting error.

\section{Validity of the correction of the lattice light shifts}
We confirm the validity of the lattice light shift correction by measuring the ISs of the $4f^{13}5d6s^2 \: (J=2) \leftrightarrow 4f^{13}6s^26p_{3/2} \: (J=3)$ transition at 792 nm, for which the lattice laser is near-resonant.
In fact, by our calculation, we confirm that this transition~\cite{Ishiyama2023-tv} makes a dominant contribution in the lattice light shift, about two orders of magnitudes larger than other transitions, 
such as $\ket{g} \leftrightarrow 4f^{14}6s6p \: ^3P_1$, $4f^{14}6s6p \: ^1P_1$, $4f^{14}6s7p \: ^3P_1$, $4f^{14}6s7p \: ^1P_1$, $4f^{14}6s8p \: ^3P_1$, $4f^{14}6s8p \: ^1P_1$~\cite{Tang2018-mx}, 
and $\ket{e} \leftrightarrow 4f^{13}6s^26p_{3/2} \: (J=2)$~\cite{Ishiyama2023-tv}$, 4f^{13}6s^26p_{1/2} \: (J=3)$, $\ 4f^{13}5d6s6p \: (J=2)$.
For the last two transitions, we employ the E1 reduced matrix elements calculated in this work, which are 1.979 and 0.290 a.u., respectively.

\begin{figure}[!tb]
\centering
\includegraphics[width=0.8\linewidth]{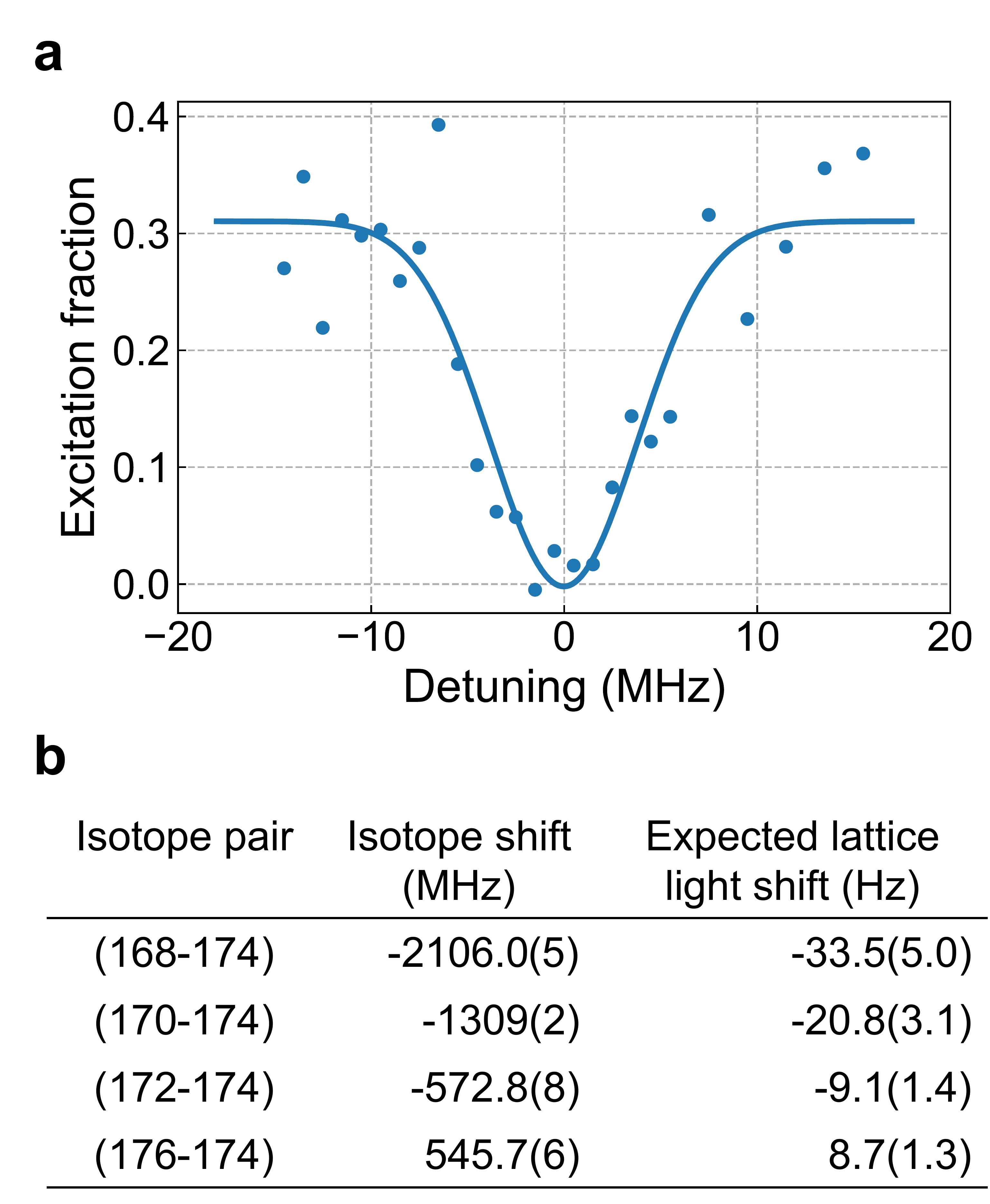}
\caption{IS measurements of the $\ket{e} \leftrightarrow 4f^{13}6s^26p_{3/2} \: (J = 3, m_J = 0)$ transition at 792 nm. \textbf{a}, A spectrum of $^{174}$Yb. The blue solid line is a Gaussian fit, yielding the FWHM of $8.9(1.1)$ MHz, where the value in the parenthesis is the $1\sigma$ fitting uncertainty. \textbf{b}, Result of the 792-nm IS measurements as well as expected lattice light shifts of the 431-nm ISs. The values in the parentheses are the $1\sigma$ statistical uncertainties derived from the fit.}
\label{fig:792IS}
\end{figure}

We perform the spectroscopy of the $\ket{e} \leftrightarrow 4f^{13}6s^26p_{3/2} \: (J=3, m_J=0)$ transition as follows.
After loading atoms into the 3D isotropic optical lattice, we simultaneously apply both the 431-nm clock laser resonant to the $\ket{g} \leftrightarrow \ket{e}$ transition and the frequency-tunable 792 nm laser.
While atoms are excited from the $\ket{g}$ to $\ket{e}$ state when the 792-nm laser is far off-resonant, the excitation fraction is reduced if the 792-nm laser is near-resonant and thus causes a sizable AC Stark shift for the $\ket{e}$ state.
As Fig.~\ref{fig:792IS}a shows, we obtain the spectrum with a linewidth of less than 10~MHz.
We perform the spectroscopy for five bosonic isotopes and determine the ISs with uncertainties less than 2~MHz, as shown in Fig.~\ref{fig:792IS}b.
The lattice light shifts are estimated by multiplying the measured ISs and the lattice light shift slopes in Fig.~\ref{fig: magic}d.
The results are displayed in Fig.~\ref{fig:792IS}b, showing consistency with the values in Fig.~3c in the main text within the $1\sigma$ uncertainties.

%
%
\section{\label{atomic calculation} Atomic calculations for electronic factors}  

\subsection{Theory}
\subsubsection{Electronic factors for field shifts}
For the expression of the field shift, most of the earlier studies (e.g., \cite{Seltzer1969PR,Bauche1976_FS,Torbohm1985PRA_FS,King1984_IS,Papoulia2016PRA_Sel,Sahoo2024JPB}) do not incorporate the isotope effect on the electronic waverunction except for Eq.~(24) of Ref.~\cite{Blundell1987-wn}. Taking the isotope-dependence for the electronic wavefunction into account, the field shift for the transition $i$ may be expressed by the difference between the expectation values of the electron-nucleus potential $V_{X}$
\be\label{eq:nu_rhoV}
\delta\nu_{\mr{FS},i}^{A'A} = \int \dd\bos{r}_\mr{e}\delta\rho_{i,A'}(\bos{r}_\mr{e})V_{A'}(\bos{r}_\mr{e}) - \int \dd\bos{r}_\mr{e}\delta\rho_{i,A}(\bos{r}_\mr{e})V_A(\bos{r}_\mr{e}), 
\ee
where
\be
\delta\rho_{i,X} = \rho_{\mr{u},X} - \rho_{\mr{l},X}; \quad \int \dd\bos{r}_\mr{e}\rho_{\mr{\alpha},X}(\bos{r}_\mr{e}) = n_\mr{e}, \quad (\alpha = \mr{u,l}).
\ee
$\rho_{\alpha,X}(\bos{r}_\mr{e})$ is the electron density of the state $\alpha$ for the mass number $X$. 
The subscripts u and l refer to the upper and lower electronic states, respectively.
$n_\mr{e}$ is the number of electrons of the system. The electron-nucleus potential can be expressed by 
\be
 V_{X}(\bos{r}_\mr{e})=-\frac{Ze^2}{4\pi\varepsilon_0}\int d\bos{r} \frac{ \rho_{\mr{n},X}\left(\bos{r}\right)}{\left|\bos{r}_\mr{e}-\bos{r}\right|}, 
\ee
where $\rho_\mr{n}$ is the normalized nuclear charge density, $e$ is the elementary charge, and $\varepsilon_0$ is the permittivity of vacuum.

When the nuclear charge distribution is spherical, the contribution from the nuclear region to the field shift can be expressed by Seltzer moment~\cite{Seltzer1969PR,Blundell1987-wn}. The following expression was explicitly derived in Ref.~\cite{Mikami2017-oh}:
%
\bea\label{eq:Seltzer_moment}
&&\int  \dd\bos{r}_\mr{e} \delta\rho_{i,X}(\bos{r}_\mr{e})V_{X}(\bos{r}_\mr{e}) 
\sim
\frac{Ze^2}{4\pi\varepsilon_0} \sum_{k=0} \frac{\xi^{(k)}_{i,X} \braket{r^{k+2}}^{X}}{(k+3)(k +2)}  \\ \nonumber
&+& \int_{r_\mr{c}}^{\infty} \mathrm{d} r_\mr{e} \int_{0}^{\pi} \sin\theta \dd \theta\int_{0}^{2\pi} \dd \phi  \left(\Xi_{i,X}(\bos{r}_\mr{e})-\sum_{k=0} \xi^{(k)}_{i,X} r^{k}\right)
,
\eea
where $\braket{r^m}^{X}$ is the $m$th nuclear charge moment. The first term represents the contribution from the inside of the nucleus, and the second term corresponds to outside the nucleus. 
$\Xi_i(\bos{r}_\mr{e})$ refers to $\rho_i(\bos{r}_\mr{e})$ outside the nucleus ($r_\mr{e}\geq r_\mr{c}$).
%
$\xi^{(k)}_{i,X}$ is the coefficient of the polynomial expansion of $\delta\rho_{i,X}(\bos{r}_\mr{e})$ at the origin
\be\label{eq:rho_expansion}
\delta\rho_{i,X}(\bos{r}_\mr{e}) \sim \delta\rho_{i,X}(r_\mr{e}) = \xi_{i,X}^{(0)}+\xi_{i,X}^{(2)} r_\mr{e}^2+\cdots,
\ee
where $r_\mr{e}$ is the distance between an electron and the center of the nucleus. Although the angular part is also required to describe $\delta\rho_{i,X}(\bos{r}_\mr{e})$ in general, the above expansion would be sufficient in the region close to the nucleus where $s$-type components (the large component of $s_{1/2}$ and the small component of $p_{1/2}$) dominantly contributes.
Although the second term of Eq.~(\ref{eq:Seltzer_moment}) also depends on $X$, we consider only the first term since the dominant contribution from the isotope shift would come from the nuclear region. 
The contributions from the electric wavefunctions in the first and second terms of Eq.~(\ref{eq:rho_expansion}) are the electronic factors for the field shift and the next-leading-order Seltzer moment, respectively
\be\label{eq:F_G4}
F_{i,X} = \frac{Z e^2}{4 \pi \varepsilon_0} \frac{\xi_{i, X}^{(0)}}{6}, \quad
G^{(4)}_{i,X} = \frac{Z e^2}{4 \pi \varepsilon_0} \frac{\xi_{i, X}^{(2)}}{9}.
\ee
We can see that the electron density at the origin is included in $F_i$, and thus only $s_{1/2}$ and $p_{1/2}$ orbitals directly contribute to $F_i$.  
The Taylor expansion of $F_{i,X}$ at the reference isotope $X=A_0$ can be expressed by 
\be\label{eq:F_taylor}
F_{i,X} = F_{i,A_0} + 2G_{i,A_0}^{(2)} 
\left( \braket{r^2}^{X}  - \braket{r^2}^{A_0}   \right) + \cdots, 
\ee
where
\be
G_{i,A_0}^{(2)}=\left.\frac{1}{2}\frac{\partial F_{i,X}}{\partial \braket{r_{X}^2}} \right|_{X=A_0}. 
\ee
By performing these expansions for both isotopes $A$ and $A'$ in Eq.~(\ref{eq:nu_rhoV}), we can obtain the following expression for the field shift 
\bea\label{eq:F_G2_G4}
\delta \nu_{\mr{FS},i}^{A^{\prime} A} 
&=& (F_{i,A_0} + 2G_{i,A_0}^{(2)}\left\langle r^2\right\rangle^{A^{\prime} A}_+)\delta\left\langle r^2\right\rangle^{A^{\prime} A}  \\ \nonumber
&+& G_{i,A_0}^{(4)} \delta\left\langle r^4\right\rangle^{A^{\prime} A} 
+ G_{i,A_0}^{(2)}\left[\delta\left\langle r^2\right\rangle^2\right]^{A^{\prime} A} + \cdots,
\eea
where $G_{i,A_0}^{(4)}$ is the first term of the Taylor expansion of $G_{i,X}^{(4)}$ at $A_0$, which is obtained in the same manner as Eq.~(\ref{eq:F_taylor}). 
The nuclear part can be expressed by
\bea
\delta\left\langle r^m\right\rangle^{A^{\prime} A}&=&\left\langle r^m\right\rangle^{A^{\prime}}-\left\langle r^m\right\rangle^A, \\ 
\left\langle r^2\right\rangle^{A^{\prime} A}_+&=&\left\langle r^2\right\rangle^{A^{\prime}}+\left\langle r^2\right\rangle^A, \\ 
\left[\delta\left\langle r^2\right\rangle^2\right]^{A'A} 
&=&  
\left(\delta\left\langle r^2\right\rangle^{A^{\prime} A_0}\right)^2-\left(\delta\left\langle r^2\right\rangle^{A A_0}\right)^2.
\eea
As shown in Eq.~(\ref{eq:F_G2_G4}), the electronic factors depend on $A_0$ but do not depend on $A$ and $A'$. 
When $\delta\nu_{\mr{FS},i}^{A'A}$ is expressed only by Seltzer moment [the first term of Eq.~(\ref{eq:Seltzer_moment})], 
$G_{i}^{(2)}$ and the reference isotope will not appear in the formula. 
In this case, however, the difference in the nuclear charge moment (e.g., $\delta\left\langle r^2\right\rangle^{A^{\prime} A}$) cannot be extracted because the electronic factors are different (e.g., $F_{i,A'}\ne F_{i,A}$). 
When taking the ratio of $F_{i, A_0}$ and $F_{j, A_0}$, the term of $(2G_{i, A_0}^{(2)}\left\langle r^2\right\rangle_{+}^{A^{\prime} A})$ would be negligible because the ratio of $G_{i, A_0}^{(2)}/F_{i, A_0}\sim0.29$ fm$^{-2}$ is almost constant from Tables \ref{tbl:F_theory} and \ref{tbl:G2_theory}, and the isotope-pair dependence of $\left\langle r^2\right\rangle_{+}^{A^{\prime} A}$ would be small. 
In our calculations, $A_0=174$ was employed.


\subsubsection{Property calculation and implementation}
We newly implemented the one-electron integral for the Yukawa-type potential in the DIRAC code. In the point-charge approximation, the interaction potential between an electron and a neutron for the particle shift term is expressed by
\begin{equation}\label{eq:Yukawa}
V_\mr{PS}(r_\mr{e}) = \alpha_\mr{NP} V_\mr{Y}(r_\mr{e}); \quad V_\mr{Y}(r_\mr{e})=\hbar c\frac{e^{-m_{\phi}cr_\mr{e}/\hbar}}{r_\mr{e}}. 
\end{equation}
Here, $\alpha_{\mathrm{NP}}$ is defined as $(-1)^{1+s_\phi} y_\mathrm{e} y_\mathrm{n} / (4 \pi \hbar c)$. $s_\phi$ and $m_{\phi}$ are the spin and mass of the new particle, respectively.
$y_{e(n)}$ represents the coupling constant between the new particle and an electron (neutron).
The corresponding electronic factor $X_i$ is defined by
\be
X_i = \braket{\Psi_\mr{u}|V_\mr{Y}|\Psi_\mr{u}} - \braket{\Psi_\mr{l}|V_\mr{Y}|\Psi_\mr{l}},
\ee
where the index $i$ refers to the transition between the upper electronic state $\ket{\Psi_\mr{u}}$ and the lower state $\ket{\Psi_\mr{l}}$. 
The numerical integration module in the DIRAC code~\cite{Sunaga2022JCP} is employed for the one-electron integration of $V_\mr{Y}$. The numerical radial grid implemented in the GRASP code~\cite{dyall1989grasp} was employed for the numerical integration. 

%
For the property calculations at the equation-of-motion coupled-cluster (EOM-CC) methods, we employed the finite-field perturbation method~\cite{Pople1968JCP_FFPT,Norman2018_property,Kuroda2022PRA}. In this approach, we add the operator for the target property ($\hat{O}$) to the electronic Hamiltonian, $\hat{H}_0$,
\be\label{eq:FFPT_H+O}
\hat{H}(\lambda)=\hat{H}_0+\lambda\hat{O},
\ee
where $\lambda$ is the perturbation strength.
Using the associated energy for the above Hamiltonian ($E(\lambda)$), the expectation value of the property can be expressed as follows:
\be
\braket{\hat{O}}=\left.\frac{d E(\lambda)}{d \lambda}\right|_{\lambda=0}.
\ee
We employed the six-point central difference formula~\cite{Fornberg1988MC_FFPT,Knecht2011TCA_Mossbauer} for the differentiation
\be\label{eq:six_point}
\braket{\hat{O}} \sim \frac{-E^{-3}+9 E^{-2}-45 E^{-1}+45 E^{+1}-9 E^{+2}+E^{+3}}{60 \lambda^{\text {opt }}},
\ee
where $E^n$ refers to $E(n\lambda^\mr{opt})$. 
To obtain the difference between the electronic factors at the ground and excited states, we employed the excitation energy obtained with the EOM-CC calculation as $E^n$. It is known that the contribution from the Hartree-Fock (HF) energy should be analytically obtained \cite{Knecht2011TCA_Mossbauer,Pernpointner2001JCP_FFPT}. In this case, however, the HF contributions are canceled in the coupled cluster singles and doubles (CCSD) and EOM-CC methods since the normal-ordered Hamiltonian is employed in both theories, allowing for the separation of reference (HF) and correlation energies.

The atomic properties can be calculated in an analytical manner at the GASCI level~\cite{Knecht_thesis}. 
$F_\alpha$ ($\alpha=\mr{u},\mr{l}; F_i=F_\mr{u}-F_\mr{l}$) was calculated using the Fermi-contact-type one-electron integral, which is implemented in the DIRAC code. $G^{(2)}_\alpha$ was calculated based on the two-point central difference formula
\bea
G_{\alpha, A_0}^{(2)} 
&\sim& \frac{1}{2}\frac{ F_{\alpha} \left(\braket{r_{A_0}^2}+h\right) - F_{\alpha} \left(\braket{r_{A_0}^2}-h\right) }{ 2h} \\ \nonumber
&=& \frac{1}{2}\frac{ F_{\alpha, A_+} - F_{\alpha, A_-} }{ 2h}, \\ \nonumber
\eea
where a step size $h \sim 3.2$ fm$^2$ was employed referring to our test. In the practical calculations, we employed $A_+=210$ and $A_-=141$. The mass number $X$ and $\braket{r_{X}^2}$ was converted based on an empirical fitting equation, $\braket{r_{X}^2}^{1/2}/\mr{fm}=0.836X^{1/3}+0.57$ \cite{Johnson_Soff_AtNucDatTab1985}.


%
%
\subsection{\label{comp_det} Computational detail}
We employed the development versions of the DIRAC code (githash: 507584eb6 and bb8403e)~\cite{saue2020dirac,DIRAC22} for the relativistic electronic structure calculation, otherwise explicitly mentioned.
For the electron correlation calculations, the \PP\ and \DD\ states were calculated with the equation-of-motion coupled-cluster (EOM-CC) singles and doubles methods using the EOMCC \cite{shee2018equation} and RELCCSD \cite{Visscher_Lee_Dyall_1996} modules. 
In the EOM-CC calculation, we employed the electronic excitation
(EOM-CC-EE) model for \PP\ and \Dtw, and the electron attachment (EOM-CC-EA) model for \DD\ and \Dth, respectively. For the HF and CCSD calculations, the electron configuration $4f^{14}6s^{2}$ ($4f^{14}6s^{0}$) was employed for the reference state of the EOM-CC-EE (EOM-CC-EA) calculations. We turned on the \texttt{.OVERLAP} option in the EOM-CC calculation to obtain the target states stably and to avoid the inner-electron excitation states except for the calculation for $X_i$ of the \DD\ and \Dth\ states. This option allows the sorting of the left/right-hand side eigenvalues and eigenvectors in descending order of overlap between the initial and generated trial vectors \cite{Zuev2015JCC_EOM_overlap,shee2018equation}.
We employed the three relativistic Hamiltonians: Dirac-Coulomb-Gaunt (DCG), Dirac-Coulomb (DC), and the exact two-component molecular mean-field approximation Hamiltonian (X2Cmmf~\cite{Sikkema_Visscher_Saue_Ilias_2009}) based on the DCG Hamiltonian, which we denote as $^2$DCG$^M$. The Gaunt term was added only at the HF level. The integrals for the small components (SS$\mid$SS) were explicitly included.
The Gaussian-type nuclear charge distribution model \cite{visscher1997ADNDT} was used for the electron-nucleus potential. 
The GRASP code \cite{dyall1989grasp} was employed to investigate the effects of the nuclear models.

The generalized-active-space configuration interaction (GASCI) method~\cite{Fleig2003JCP,Knecht2010JCP} was employed for the $\g\: \leftrightarrow \:$\JJ\ and $^2S_{1/2} \leftrightarrow \:$\FF\ transitions. 
The atomic orbitals are obtained with the HF method for the closed-shell system, and the average-of-configuration HF (AOC-HF) method \cite{Thyssen_thesis} for open-shell systems. The employed electron configurations are as follows: $4f^{14}6s^2$ for $\g$, $4f^{13}5d^16s^2$ for \JJ, $4f^{14}6s^1$ for $^2S_{1/2}$, and $4f^{13}6s^2$ for \FF\ states. In this notation, thirteen electrons are distributed in $4f$ orbitals, one electron is distributed in $5d$ orbitals, and $6s$ orbitals are treated as a closed shell in the configuration $4f^{13}6s^25d^1$. 
Our correlation model of the GASCI calculation is labeled as kk(1)\_mm(2)\_nn(3), where one, two, or three holes are allowed in the orbitals indicated by kk, mm, and nn. 
In addition, we imposed one hole in the $4f$ orbitals for the \JJ\ and \FF\ states, and we included one electron in the $5d$ orbitals for the \JJ\ state. An example for the 4f4s5s5p(1)\_6s(5d)(2) model is shown in Table \ref{tbl:CI_example}, where (5d) implies that the occupation in the $5d$ orbital is allowed only for \JJ\ state, and $5d$ orbitals are included in the virtual space in other states. The contributions from the frozen orbitals are included at the (AOC-)HF level.
%
%
\begin{table}[htbp!]
\centering
\caption{
Example for generalized active space models for the CI wave functions 
in the case of the 4f4s4p5s5p(1)\_6s(5d)(2) model with the ae3z+3s2p basis set and the cutoff energy of 30 $E_\mr{h}$. 
}\label{tbl:CI_example}
\begin{tabular}{@{\extracolsep{\fill}}cc cc cc cc cc@{}}
\hline\hline
 & Number of & \multicolumn{8}{c}{accumulated \# of electrons}\tabularnewline
 &  Kramers  & \multicolumn{2}{c}{$^1S_0$} & \multicolumn{2}{c}{$4f^{13}5d6s^{2}$} & \multicolumn{2}{c}{$^2S_{1/2}$} & \multicolumn{2}{c}{$4f^{13}6s^{2}$}\tabularnewline
Orbital &  pairs & min & max & min & max & min & max & min & max\tabularnewline
\hline
virtual & 112 & 26 & 26 & 26 & 26 & 25 & 25 & 25 & 25\tabularnewline
5d & 5 & 26 & 26 & 24 & 26 & 25 & 25 & 25 & 25\tabularnewline
6s & 1 & 24 & 26 & 23 & 25 & 23 & 25 & 23 & 25\tabularnewline
4s5s5p & 5 & 23 & 24 & 22 & 23 & 23 & 24 & 22 & 23\tabularnewline
4f & 7 & 13 & 14 & 12 & 13 & 13 & 14 & 12 & 13\tabularnewline
frozen & (22) &  &  &  &  &  &  &  & \tabularnewline
\hline\hline
\end{tabular}
\end{table}

The relativistic dyall basis sets \cite{Dyall_4f} were employed in the uncontracted form. The small components of the basis sets were generated by the restricted kinetic balance \cite{Stanton1984JCP,Dyall1984JPB}, which is the default of the DIRAC code \cite{saue2020dirac}. For the electronic factors ($F_i$ and \Gtwo) calculation, we employed the dyall.ae3z and dyall.ae4z basis sets, including the tight exponents shown in Table~\ref{tbl:tight_exponent}, which we note as ae3z+3s2p and ae4z+3s2p. 
%
%
\begin{table*}[htbp!]
\caption{
Tightest and added exponents, and average relative error from the numerical wavefunction obtained using the simplex procedure \cite{sunaga2021MP}.
}\label{tbl:tight_exponent}
\begin{tabular}{l |cc|cc}
\hline\hline
 & \multicolumn{2}{c|}{dyall.4zp}  & \multicolumn{2}{c}{dyall.3zp} \tabularnewline
 & $s$ & $p$ & $s$ & $p$\tabularnewline
 \hline
tightest & 6.5973426$\times10^7$ & 4.9178921$\times10^7$ & 6.5666750$\times10^7$ & 1.9107953$\times10^7$\tabularnewline
added & 2.4788468$\times10^8$ & 2.1132889$\times10^8$ & 2.2575015$\times10^8$ & 2.5777541$\times10^7$\tabularnewline
 & 6.5778342$\times10^8$ & 5.7256503$\times10^8$ & 1.1663319$\times10^9$ & 5.6036488$\times10^7$\tabularnewline
 & 1.2942181$\times10^9$ &                       & 6.5731480$\times10^9$ &                       \tabularnewline
Error (\%) & 1.0$\times10^{-4}$ & 4.8$\times10^{-4}$ & 9.4$\times10^{-3}$ & 5.9$\times10^{-3}$\tabularnewline
 \hline\hline
\end{tabular}
\end{table*}
These tight exponents were optimized for the Fermi-contact-type integral (i.e., the density at the initial grid point) of the numerical wavefunction with the $6s^16p_{1/2}^1$ electronic configuration obtained with the GRASP code.
The optimization was performed using the dyall.$n$zp ($n=3,4$) basis set and the generated exponents were added to the corresponding dyall.ae$n$z basis sets. The optimization was performed for $s$ and $p$ exponents independently.  

For the calculation of $X_i$, we employed the EOM-CC method based on the ${ }^2 \mathrm{DCG}^M$ Hamiltonian for \PP, \DD, \Dth, and \Dtw. The electrons higher than --30 $E_\mr{h}$ (i.e., higher than $3p$ orbital) were correlated, and the virtual space was truncated 100 $E_\mr{h}$. For the \JJ\ and \FF\ states, we employed the 5p(1)\_4f6s(5d)(2) correlation model with the virtual truncation of 30 $E_\mr{h}$ based on the DCG Hamiltonian. The contributions from the inner orbitals (from $1s$ to $5s$ orbitals) were included at the (AOC-)HF level based on the DCG Hamiltonian. For the calculations of $F_i$ and \Gtwo, we employed various computational methodologies, as listed in Tables \ref{tbl:comp_det_578_411} and \ref{tbl:comp_det_431_467}.
The dyall.v3z basis set was employed for \JJ\ and \FF\ states. The dyall.v4z was employed for \PP\ and \Dtw, and the dyall.v3z with even-temporally augmented $s, p, d$, and $f$ exponents was employed for \DD\ and \Dth\ to obtain the numerical derivative in a numerically stable manner.



%
%
\begin{table*}[htbp!]
\caption{
Summary of $F_i$ ($\mathrm{GHz} / \mathrm{fm}^2$) and \Gtwo\ ($\mathrm{MHz} / \mathrm{fm}^4$) for the $i=578$ (\PP) and $411$ (\DD) transitions at various calculation levels, which are used for obtaining the corrections in Tables \ref{tbl:F_theory} and \ref{tbl:G2_theory}. 
The virtual cutoff energy (virtual) is in $E_\mr{h}$.
The calculation conditions for the blank rows are the same as those for the rows immediately above them.
The correction terms written on only one line were obtained taking the difference from the baseline.
}\label{tbl:comp_det_578_411}
\begin{tabular}{llllcl rccc}
\hline\hline
 & Hamiltonian & basis & method & virtual & correlated elec. & $F_{578}$ & $G^{(2)}_{578}$ & $F_{411}$ & $G^{(2)}_{411}$ \tabularnewline
 \hline
baseline & DC & ae3z+3s2p & EOM-CC & 100 & all & $-$10.448 & 29.349 & $-$17.213 & 47.521\tabularnewline\rule[0mm]{0mm}{5mm}
$\Delta$Basis & DC & {ae4z+3s2p} & EOM-CC & 100 & all & $-$10.553 & 29.478 & $-$17.348 & 48.190\tabularnewline \rule[0mm]{0mm}{5mm}
$\Delta$Triple & DCG & ae3z+3s2p & {CCSD(T)} & 100 & higher than 3p & $-$15.019\footnotemark[1] & 41.922\footnotemark[1] & $-$17.078 & 47.138\tabularnewline
 &  &  & {EOM-CC} &  &  & $-$15.149\footnotemark[1] & 42.264\footnotemark[1] & $-$17.101 & 46.782\tabularnewline\rule[0mm]{0mm}{5mm}
$\Delta$Virt. cutoff & DC & ae3z+3s2p & EOM-CC  & {3000} & higher than 3p & $-$10.461 & 29.485 & $-$17.227 & 47.593\tabularnewline \rule[0mm]{0mm}{5mm}
$\Delta$Gaunt & {DCG} & ae4z+3s2p & HF &  &  & $-$9.324 & 26.049 & -- & --\tabularnewline
 & {DC} &  &  &  &  & $-$9.274 & 25.907 & -- & --\tabularnewline\rule[0mm]{0mm}{5mm}
$\Delta$Gaunt & {DCG} & ae3z+3s2p & EOM-CC & 100 & higher than 3p & -- & -- & $-$17.101 & 46.782\tabularnewline
 & {DC} &  &  &  &  & -- & -- &  $-$17.177 & 47.436\tabularnewline
 \hline\hline
\end{tabular}
\footnotetext[1]{These are the values of the $^2P_{1/2}$ state. See text.}
\end{table*}
%
%
\begin{table*}[htbp!]
\caption{
Summary of $F_i$ ($\mathrm{GHz} / \mathrm{fm}^2$) and \Gtwo\ ($\mathrm{MHz} / \mathrm{fm}^4$) for $i=431$ [\JJ] and $467$ (\FF) transitions at various calculation levels, which are used for obtaining the corrections in Tables \ref{tbl:F_theory} and \ref{tbl:G2_theory}.
The virtual cutoff energy (virtual) is in $E_\mr{h}$.
The contributions from the frozen orbitals are not included in `only valence' values. 
The calculation conditions for the blank rows are the same as those for the rows immediately above them.
The correction terms written on only one line were obtained, taking the difference from the DCG/4f(1)\_5s5p6s(5d)(2)/ae3z+3s2p level. $\Delta$Gaunt is calculated as the difference from the baseline.
The values of $\Delta$Inner corr.(4p) and $\Delta$Inner corr.(3s) are obtained by taking the difference from the value at the 4f4s5s5p(1)\_6s(5d)(2) level.
}\label{tbl:comp_det_431_467}
\begin{tabular}{lll cl cccc}
\hline\hline
 & Hamiltonian & basis & virtual & CI model & $F_{431}$ & $G^{(2)}_{431}$ & $F_{467}$ & $G^{(2)}_{467}$\tabularnewline
 \hline
only valence & DC & ae3z+3s2p & 30 & 4f(1)\_5s5p6s(5d)(2) & 18.194 & --49.640 & 37.794 & --103.60\tabularnewline
baseline  & DC & ae3z+3s2p & 30 & 4f(1)\_5s5p6s(5d)(2) & 20.123 & --54.911 & 40.272 & --110.40\tabularnewline \rule[0mm]{0mm}{5mm}
$\Delta$Basis & DCG & {ae4z+3s2p} & 10 & 4f(1)\_5s5p6s(5d)(2) & 20.341 & --56.089 & 40.539 & --112.10\tabularnewline
 &  & {ae3z+3s2p} &  &  & 20.441 & --55.782 & 40.575 & --111.23\tabularnewline \rule[0mm]{0mm}{5mm}
$\Delta$Triple & DCG & ae3z+3s2p & 10 & 4f5s(1)\_6s(5d){(3)} & 19.819 & --54.335 & -- & --\tabularnewline
 &  &  &  & 4f5s(1)\_6s(5d){(2)} & 19.798 & --54.282 & -- & --\tabularnewline \rule[0mm]{0mm}{5mm}
$\Delta$Triple & DCG & ae3z+3s2p & 30 & 4f5s5p(1)\_6s{(3)} & -- & -- & 42.118\footnotemark[1] & --115.51\footnotemark[1]\tabularnewline
 &  &  &  & 4f5s5p(1)\_6s{(2)} & -- & -- & 42.372 & --116.19 \tabularnewline \rule[0mm]{0mm}{5mm}
$\Delta$Virt. cutoff & DCG & ae3z+3s2p &{50}& 4f(1)\_5s5p6s(5d)(2) & 20.206 & --55.134 & 40.457 & --110.90\tabularnewline \rule[0mm]{0mm}{5mm}
 %
$\Delta$Gaunt & {DCG} & ae3z+3s2p & 30 & 4f(1)\_5s5p6s(5d)(2) & 20.216 & --55.163 & 40.372 & --110.67\tabularnewline \rule[0mm]{0mm}{5mm}
 %
$\Delta$Inner corr.(4s) & DCG & ae3z+3s2p & 30 & 4f{4s}5s5p(1)\_6s(5d)(2) & 19.160 & --52.169 & 41.703 & --114.25\tabularnewline
 &  &  &  & 4f5s5p(1)\_6s(5d)(2) & 19.480 & --53.145 & 42.372 & --116.19\tabularnewline
$\Delta$Inner corr.(4p) & DCG & ae3z+3s2p & 30 & 4f{4s4p}5s5p(1)\_6s(5d)(2) & 19.096 & --51.879 & 41.679 & --114.09\tabularnewline
$\Delta$Inner corr.(3s) & DCG & ae3z+3s2p & 30 & 4f{3s4s}5s5p(1)\_6s(5d)(2) & 19.000 & --51.742 & 41.286 & --113.11\tabularnewline
\hline\hline
\end{tabular}
\footnotetext[1]{The 4f5s5p(1)\_6s(2) model were employed for $^2S_{1/2}$.}
\end{table*}

For numerical differentiation provided in Eq.~(\ref{eq:six_point}), the following step sizes were employed for $F_{i}$: $\lambda_\mathrm{opt} = 1.0\times10^{-8}$ and $2.0\times10^{-8}$ for cations (\DD\ and $^2P_{1/2}$) and \PP, respectively. For $X_i$, the employed step sizes were $\lambda_\mathrm{opt} = 1.0\times10^{-5}$ for $1.0 \leq m_{\phi} \leq 1.0\times10^6$, $\lambda_\mathrm{opt} = 1.0\times10^{-3}$ for $1.0\times10^6 < m_{\phi} \leq 1.0\times10^7$, and $\lambda_\mathrm{opt} = 0.1$ for $1.0\times10^7 < m_{\phi} \leq 1.0\times10^8$. $\lambda_\mathrm{opt}$ is in a.u. and $m_{\phi}$ is in eV.

The nuclear model effect ($\Delta$Fermi) was obtained at the AOC-HF level using the GRASP code \cite{dyall1989grasp}. Since the density at the origin cannot be taken in the calculation using numerical grid points, we used the density at the initial grid point ($r\sim1.4\times10^{-8} \; a_0$, where $a_0$ refers to the Bohr radius).
The electron configurations employed for \PP, \DD, \JJ, and \FF\ are $4f^{14}6s6p_{1/2},4f^{14}5d_{5/2},4f^{13}6s5d$, and $4f^{13}6s^2$, respectively.
The relative corrections (in percentage) were obtained for $F_{X,i}$, where $X= 141,174,$ and $210$, using the Gauss- and Fermi-type models. The values of $\Delta$Fermi listed in Tables \ref{tbl:F_theory} and \ref{tbl:G2_theory} were obtained from the relative corrections and the values of baseline.

\subsection{Calculation results}
The calculated values of $F_i$ and $G^{(2)}_i$ are listed in Tables \ref{tbl:comp_det_578_411} and \ref{tbl:comp_det_431_467}. 
The corrections and uncertainty due to the basis set ($\Delta$Basis), high-order electron correlation effects ($\Delta$Triple), truncation of virtual space ($\Delta$Virt. cutoff), correlations of inner electrons ($\Delta$Inner corr.), high-order relativistic effects ($\Delta$Gaunt) are obtained by changing the corresponding computational conditions. 
We approximately estimated the higher-order correlation effects ($\Delta$Triple) for the $i=578$ transition using the $^2S_{1/2}\: \leftrightarrow \:^2P_{1/2}$ transition and its ratio, where the single reference treatment based on the $6s^06p^1_{1/2}$ configuration. The reason for this is that the single reference CCSD(T) method may not work well for the electronic configuration of $6s^16p^1_{1/2}$.  
$\Delta$Gaunt of the $i=578$ transition was obtained at the HF level because we were faced with the numerical instability at the EOM-CC level.

The corrections, uncertainties, and final values of $F_i$ and \Gtwo\ are summarized in Tables \ref{tbl:F_theory} and \ref{tbl:G2_theory}. The total uncertainty is obtained by taking the Euclidean norm of all corrections, as these corrections are assumed to be independent. 
The largest correction (uncertainty) is the correlation effects of inner electrons for the $i=431$ and 467 transitions. It implies that the inner electrons should be ideally included in the active space, although the relaxation effects due to the excitation of the $4f$ electron are included in the state-specific calculation in this study. 
The other corrections are not very substantial, which implies the computational condition for the baseline includes the dominant effects.

Table \ref{tbl:X} summarizes the calculated values of $X_i$ at $m_\phi=1$~eV and the transition frequencies $\omega_i/(2\pi)$. The values of $\omega_i/(2\pi)$ are obtained using the same computational methodologies for $X_i$, without the perturbation term $\lambda\hat{O}$ in Eq.~\ref{eq:FFPT_H+O}. The values of $X_i$ as a function of the new-boson mass are visualized in Extended Data Fig. 6 of the main text.
From a theoretical point of view, the EOM-CC theory employed in this study can include higher correlation effects that are not included in the CI theory employed in the previous works. 
%

%
%
\begin{table}[htbp!]
\caption{
Calculated values of $F_i$ ($\mathrm{GHz} / \mathrm{fm}^2$) for the four transitions including the high-order corrections.
The corrections are obtained from Tables \ref{tbl:comp_det_578_411} and \ref{tbl:comp_det_431_467}.
}\label{tbl:F_theory}
\begin{tabular}{l rr rr}
\hline 
\hline 
& 578 & 411 & 431 & 467 \tabularnewline
\hline 
(only valence) & -- & -- & (18.194) & (37.794)\tabularnewline
baseline & --10.448 & --17.213 & 20.123 & 40.272\tabularnewline
$\Delta$Basis & --0.105 & --0.135 & --0.101 & --0.036\tabularnewline
$\Delta$Triple & 0.091 & 0.023 & 0.021 & --0.254\tabularnewline
$\Delta$Virt. cutoff & --0.013 & --0.014 & --0.010 & 0.085\tabularnewline
$\Delta$Inner corr.(4s) & -- & -- & --0.319 & --0.669\tabularnewline
$\Delta$Inner corr.(4p) &--  & -- & --0.065 & --0.024\tabularnewline
$\Delta$Inner corr.(3s) & -- & -- & --0.160 & --0.417\tabularnewline
$\Delta$Gaunt & --0.050 & 0.076 & 0.094 & 0.099\tabularnewline
$\Delta$Fermi & 0.169 & 0.278 & --0.295 & --0.617\tabularnewline
Final result & --10.357 & --16.985 & 19.287 & 38.439\tabularnewline 
uncertainty  & 0.225  & 0.319 & 0.488 & 1.042 \tabularnewline
\hline
\hline
\end{tabular}
\end{table}

%
%
\begin{table}[htbp!]
\caption{
Calculated values of \Gtwo\ ($\mathrm{MHz} / \mathrm{fm}^4$) for the four transitions including the high-order corrections.
The corrections are obtained from Tables \ref{tbl:comp_det_578_411} and \ref{tbl:comp_det_431_467}.
}\label{tbl:G2_theory}
\begin{tabular}{l rr rr}
\hline \hline 
& 578 & 411 & 431 & 467 \tabularnewline
 \hline 
(only valence) & -- & -- & (--49.640) & (--103.60)\tabularnewline
baseline & 29.349 & 47.521 & --54.911 & --110.40\tabularnewline
$\Delta$Basis & 0.130 & 0.669 & --0.307 & --0.87\tabularnewline
$\Delta$Triple & --0.240 & 0.357 & --0.053 & 0.68\tabularnewline
$\Delta$Virt. cutoff & 0.136 & 0.071 & 0.029 & --0.23\tabularnewline
$\Delta$Inner corr.(4s) & -- &--  & 0.976 & 1.94\tabularnewline
$\Delta$Inner corr.(4p) &--  & -- & 0.290 & 0.16\tabularnewline
$\Delta$Inner corr.(3s) &  --&--  & 0.427 & 1.14\tabularnewline
$\Delta$Gaunt & 0.141 & --0.654 & --0.252 & --0.27\tabularnewline
$\Delta$Fermi & 0.377 & 0.627 & --0.668 & --1.40\tabularnewline
Final result & 29.894 & 48.591 & --54.469 & --109.25\tabularnewline
uncertainty  & 0.505  & 1.183 & 1.352 & 2.90 \tabularnewline
\hline \hline 
\end{tabular}
\end{table}

%
%
\begin{table}[htbp!]
\caption{
Calculated values of $X_i$ at $m_{\phi}=1$ eV and transition frequencies $\omega_i /(2 \pi)$ from this work (TW) and literature for the six transitions. 
Experimental values are taken from Ref.~\cite{NIST_ASD}.
}\label{tbl:X}
\begin{tabular}{@{} l S[table-format=4.4] S[table-format=4.4] S[table-format=1.4]S[table-format=4.4] S[table-format=1.4]S[table-format=1.4] @{}}
\hline\hline
 & {578} & {467} & {436} & {431} & {411} & {361} \tabularnewline 
 \hline
  & \multicolumn{6}{c}{$X_i$~($\times 10^3$ THz)} \tabularnewline 
TW & -42.333 & -293.02 & 46.749 & -206.53 & 40.545 & 9.3176\tabularnewline
\cite{Kawasaki2024-co} & -39.49 & -295.06 & 49.74 & -194.34 & 43.370 & \tabularnewline
\cite{Hur2022PRL}\footnotemark[1] & -55.729 & -730.40 & 48.419 &  & 44.145 & \tabularnewline
\cite{Hur2022PRL}\footnotemark[2] & -42.855 & -352.38 & 48.634 &  & 43.158 & 5.6683 \\[1mm]
  & \multicolumn{6}{c}{$\omega_i /(2 \pi)$ ~(THz)} \tabularnewline
TW & 509.88 & 358.64 & 720.42 & 332.97 & 756.75 & 856.17\tabularnewline
\cite{Kawasaki2024-co}  & 500.53 & 727.06 & 649.86 & 810.67 & 691.88 & 875.19\tabularnewline
\cite{Hur2022PRL}\footnotemark[1] & 458.36 & 580.12 & 770.13 &  & 808.11 & \tabularnewline
\cite{Hur2022PRL}\footnotemark[2] & 522.78 & 1051.44 & 679.86 &  & 707.00 & 819.47\tabularnewline
exp. & 518.29 & 642.12 & 688.35 & 695.17 & 729.48 & 829.76\tabularnewline
\hline\hline
\end{tabular}
\footnotetext[1]{The values obtained using the GRASP code.}
\footnotetext[2]{The values obtained using the AMBiT code.}
\end{table}

The contributions from each atomic orbital to $F_i$ and \Gtwo, and $X_i$ for the 431 and 467 nm transitions are listed in Tables~\ref{tbl:F_G_orb} and \ref{tbl:X_orb}, respectively. Substantial contributions from orbitals that are not directly involved in the electron excitation are found. For example, the $5s_{1/2}$ orbital is one of the dominant contributions to $F_i$ and \Gtwo, and summed contributions from $1s_{1/2}$ to $5p_{3/2}$ orbitals are comparable in magnitude to the total values of $X_i$ but have the opposite sign.
Such large contributions have not been reported in previous works: for example, $^3P_0$ and $^3P_1$ states (Table XIII of Ref.~\cite{Schelfhout2021PRA}), as well as the $^2D_{3/2}$ and $^2D_{5/2}$ states and the \FF\ state without orbital relaxation effects  (Section VI Appendices C of Ref.~\cite{Hur2022thesis}). In previous calculations of the \JJ\ and \FF\ states, such orbital relaxation effects were not taken into account, as a single common set of atomic orbitals was used in the CI calculations. 

If we use the (AOC-)HF values shown in Table~\ref{tbl:F_G_orb}, which are in theory less accurate than the final values listed in Table~\ref{tbl:F_theory}, the agreement with the experimental values shown in Table~\ref{tab:2D KP} is improved. It indicates that a less accurate calculation can agree with the experiment because of the error cancellation. The development of a many-body theory that can treat the strong correlations in 4f-electron excited states is encouraged.

%

%
%
\begin{table}[htbp!]
\caption{
Orbital contributions to $F_i$ and \Gtwo\ of the 431 and 467 transitions obtained at the DCG-(AOC-)HF/ae4z\_3s2p level of theory. The employed electronic configurations are as follows: $4f^{14}6s^2$ for $\g$, $4f^{13}5d^16s^2$ for \JJ, $4f^{14}6s^1$ for $^2S_{1/2}$, and $4f^{13}6s^2$ for \FF\ states.
}\label{tbl:F_G_orb}
\begin{tabular}{c S[table-format=4.4] S[table-format=4.4] S[table-format=4.4] S[table-format=4.4]}
\hline\hline
 &  \multicolumn{2}{c}{$F_i$ (GHz/fm$^2$)} &  \multicolumn{2}{c}{\Gtwo\ (MHz/fm$^4$)} \tabularnewline
 & {431} & {467} & {431} & {467} \tabularnewline
 \hline
$1s_{1/2}$ & -0.885 & -1.148 & 2.444 & 3.174\tabularnewline
$2s_{1/2}$ & -2.113 & -1.925 & 5.854 & 5.335\tabularnewline
$2p_{1/2}$ & -0.141 & -0.149 & 0.346 & 0.364\tabularnewline
$3s_{1/2}$ & -1.058 & -0.695 & 2.939 & 1.934\tabularnewline
$3p_{1/2}$ & -0.091 & -0.083 & 0.222 & 0.203\tabularnewline
$4s_{1/2}$ & 5.822 & 6.102 & -16.113 & -16.891\tabularnewline
$4p_{1/2}$ & 0.394 & 0.376 & -0.964 & -0.921\tabularnewline
$5s_{1/2}$ & 14.386 & 18.154 & -39.903 & -50.336\tabularnewline
$5p_{1/2}$ & 0.949 & 1.155 & -2.324 & -2.829\tabularnewline
$6s_{1/2}$ & 4.327 & 19.577 & -12.055 & -54.431\tabularnewline
\hline
total & 21.591 & 41.364 & -59.553 & -114.397\tabularnewline
\hline\hline
\end{tabular}
\end{table}

%
%
\begin{table}[htbp!]
\caption{
Orbital contributions to $X_i$ ($\times 10^3$ THz) at $m_\phi=1$ eV of the 431 and 467 transitions obtained at the DCG-(AOC-)HF/dyall.v3z level of theory. The employed electronic configurations are as follows: $4f^{14}6s^2$ for $\g$, $4f^{13}5d^16s^2$ for \JJ, $4f^{14}6s^1$ for $^2S_{1/2}$, and $4f^{13}6s^2$ for \FF\ states.
}\label{tbl:X_orb}
\begin{tabular}{c S[table-format=4.4] S[table-format=4.4]}
\hline\hline
 & {431} & {467} \tabularnewline
 \hline
 $ 1s_{1/2}  $ & -0.23 & -0.29\tabularnewline
 $ 2s_{1/2}  $ & -0.97 & -0.89\tabularnewline
 $ 2p_{1/2}  $ & -0.87 & -0.88\tabularnewline
 $ 2p_{3/2}  $ & -1.64 & -1.62\tabularnewline
 $ 3s_{1/2}  $ & -0.78 & -0.53\tabularnewline
 $ 3p_{1/2}  $ & -0.86 & -0.76\tabularnewline
 $ 3p_{3/2}  $ & -1.53 & -1.34\tabularnewline
 $ 3d_{3/2}  $ & -1.75 & -1.77\tabularnewline
 $ 3d_{5/2}  $ & -2.73 & -2.75\tabularnewline
 $ 4s_{1/2}  $ & 6.98 & 7.25\tabularnewline
 $ 4p_{1/2}  $ & 8.02 & 7.66\tabularnewline
 $ 4p_{3/2}  $ & 18.06 & 17.06\tabularnewline
 $ 4d_{3/2}  $ & 26.78 & 24.67\tabularnewline
 $ 4d_{5/2}  $ & 42.68 & 39.50\tabularnewline
 $ 5s_{1/2}  $ & 36.60 & 46.46\tabularnewline
 $ 5p_{1/2}  $ & 44.75 & 54.49\tabularnewline
 $ 5p_{3/2}  $ & 93.90 & 114.06\tabularnewline
 $ 6s_{1/2}  $ & 39.98 & 351.51\tabularnewline
 $ 4f_{5/2}  $ & -447.46 & -440.58\tabularnewline
 $ 4f_{7/2}  $ & -548.80 & -539.29\tabularnewline
 $ 5d_{3/2}  $ & 179.27 & \tabularnewline
 $ 5d_{5/2} $ & 260.31 & \tabularnewline
 \hline
total & -250.30 & -328.04\tabularnewline
\hline\hline
\end{tabular}
\end{table}

\section{King plot analysis}
\subsection{Individual fit of 2D King linearity} \label{subsec: 2D}
The linearity of each 2D King plot is individually tested with the following relation:
\begin{equation}  \label{eq: individual 2D}
    \dn{2}{AA_0} = K_{12} \w{AA_0} + F_{12} \dn{1}{AA_0}.
\end{equation}
As the procedure of the statistical test, we adopt a similar method to Ref.~\cite{Ono2022-oy}.
$\chi^2$ is defined as
\begin{equation} \label{eq: individual 2D chi2}
    \chi^2 = \chi^2_{\eta} + \chi^2_{\delta \nu_1} + \chi^2_{\delta \nu_2}.
\end{equation}
Here, three terms correspond to nuclear mass ratios $\eta^A = m_A / m_{172}$~\cite{Door2025-jv}, ISs of the first and second transition, respectively:
\begin{eqnarray}
    && \chi^2_{\eta} = \sum_{A=168, 170, 174, 176} 
    \left\{ \frac{1 / \left( \w{A, A_0} - \w{172, A_0} + 1 \right) - \hat{\eta}^{A}}{\sigma_{\eta^{A}}} \right\}^2, \label{eq: mass} \nonumber \\ \\
    && \chi^2_{\delta \nu_i} = \sum_{(A',A)} 
    \left\{ \frac{ \left( \dn{i}{A'A_0} - \dn{i}{AA_0} \right) - \hat{\delta \nu}_{i}^{A'A}}{\sigma_{\dn{i}{A'A}}} \right\}^2. \label{eq: IS}
\end{eqnarray}
$\hat{(\cdot)}$ and $\sigma_{(\cdot)}$ correspond to the measured values and their experimental $1\sigma$ uncertainties, respectively, where $(A, A')$ is an isotope pair for which the IS is actually measured~\cite{Note1}.
Basically, there are 12 experimental values, since each of $\hat{\eta}^A$, $\hat{\delta \nu}_{1}^{A'A}$, and $\hat{\delta \nu}_{2}^{A'A}$ has four elements associated with four isotope pairs.
On the other hand, we define ten fitting parameters of $K_{12}$, $F_{12}$, $\w{AA_0}$, and $\dn{1}{AA_0}$, where $A = 168, 170, 172, 174$, and $A_0 = 176$.
The degree of freedom (dof) results in $(3\times4) - (2\times4+2)=2$.

We summarize best-fit parameters and nonlinearities of individual 2D King plots in Table~\ref{tab:2D KP}.
Note that, for the $361 \ (436)$ transition, six (five) isotope pairs are measured~\cite{Figueroa2022-cm, Counts2020-ca}, while the number of the minimal pairs consisting of five isotopes is four.
In Table~\ref{tab:2D KP}, we subtract $\chi^2$ and dof associated with such redundant isotope pairs, which are 2.058 (3.911) and 2 (1), respectively, for the $i=361 \ (436)$ transition. 

As for the slope of the 2D King plot $F_{12}$, the best-fit values agree with the theoretical values within a few percent except for the 431 transition. 
The larger error of the \JJ\ state would be explained by the property of the electronic configuration. 
The occupation number of the $6s$ orbital is changed through the transition of the other states, while the occupation number $6s(2)$ remains in the \JJ\ state. 
Therefore, the \JJ\ state would be more sensitive to the correlation effects, especially the relaxation effects due to the $4f$ electron excitation.

\subsection{Individual fit of 3D King linearity}
The discussion in Sec.~\ref{subsec: 2D} is straightforwardly generalized to 3D generalized King linearity 
\begin{equation} \label{eq: individual 3D}
    \dn{3}{AA_0} = k \w{AA_0} + f_1 \dn{1}{AA_0} + f_2 \dn{2}{AA_0}, 
\end{equation}
by introducing an additional $\chi^2$ term:
\begin{equation}
    \chi^2 = \chi^2_{\eta} + \chi^2_{\delta \nu_1} + \chi^2_{\delta \nu_2} + \chi^2_{\delta \nu_3}.
\end{equation}
Here, we define 15 fitting parameters of $\w{AA_0}$, $\dn{1}{AA_0}$, $\dn{2}{AA_0}$, $k$, $f_1$, and $f_2$, and the dof is basically $(4\times4) - (3\times4+3)=1$.
Best-fit parameters and nonlinearities of individual 3D King plots are summarized in Table~\ref{tab:3D KP}, where $\chi^2$ and dof associated with redundant isotope pairs are subtracted.

To obtain a constraint to the new particle at a given mass $m_\phi$, the linear relation of Eq.~(\ref{eq: individual 3D}) is replaced to 
\begin{eqnarray}  \label{eq: individual 3D PS}
    \dn{3}{AA_0} &=& k \w{AA_0} + f_1 \dn{1}{AA_0} + f_2\dn{2}{AA_0}\\
    &+& \alpha_{\mathrm{NP}} (X_3 - f_1 X_1 - f_2 X_2) (A - A_0), \nonumber
\end{eqnarray}
which is equivalent to Eq.~(18) in Methods.
Here, we add one fitting parameter $\alpha_{\mathrm{NP}}$, resulting in the dof of zero.

\subsection{Combined fit of 3D King linearities}
Here, we consider another 3D King relation of 
\begin{equation} \label{eq: combined 3D}
    \dn{4}{AA_0} = k' \w{AA_0} + f_1' \dn{1}{AA_0} + f_2' \dn{2}{AA_0},
\end{equation}
in addition to Eq.~(\ref{eq: individual 3D}) to perform the simultaneous test of their linearities.
$\chi^2$ is defined as
\begin{equation} \label{eq: combined 3D chi2}
    \chi^2 = \chi^2_{\eta} + \chi^2_{\delta \nu_1} + \chi^2_{\delta \nu_2} + \chi^2_{\delta \nu_3} + \chi^2_{\delta \nu_4}.
\end{equation}
Since King linearity parameters of $k'$, $f_1'$, and $f_2'$ are additionally defined for the fourth transition, the dof is then given by $(5\times4) - (3\times4+2\times3)=2$, which increases by one compared to an individual fit.
We obtain a $\chi^2$ minimum of $7.3\times 10^3$ for the transition sets of (578, 431, 411) and (578, 431, 467).

In setting a constraint to the new physics, besides Eq.~(\ref{eq: individual 3D PS}), the PS term is also introduced for the fourth transition:
\begin{eqnarray} \label{eq: combined 3D PS}
    \dn{4}{AA_0} &=& k' \w{AA_0} + f_1' \dn{1}{AA_0} + f_2' \dn{2}{AA_0} \\
    &+& \alpha_{\mathrm{NP}} (X_3 - f_1' X_1 - f_2' X_2) (A - A_0), \nonumber
\end{eqnarray}
Importantly, $\alpha_{\mathrm{NP}}$ in Eq.~(\ref{eq: combined 3D PS}) is common to in Eq.~(\ref{eq: individual 3D PS}).
Therefore, the dof is $(5\times4) - (3\times4+2\times3+1)=1$, allowing us to test the existence of the new particle at a given mass.
For example, the nonlinearity results in $\chi^2 = 1.3 \times 10^3$ at $m_{\phi} = 10$~eV, suggesting that the new particle is not the only source of the nonlinearity in 3D King plots.

\begin{figure}[!t]
 \centering
 \includegraphics[width=0.9\linewidth]{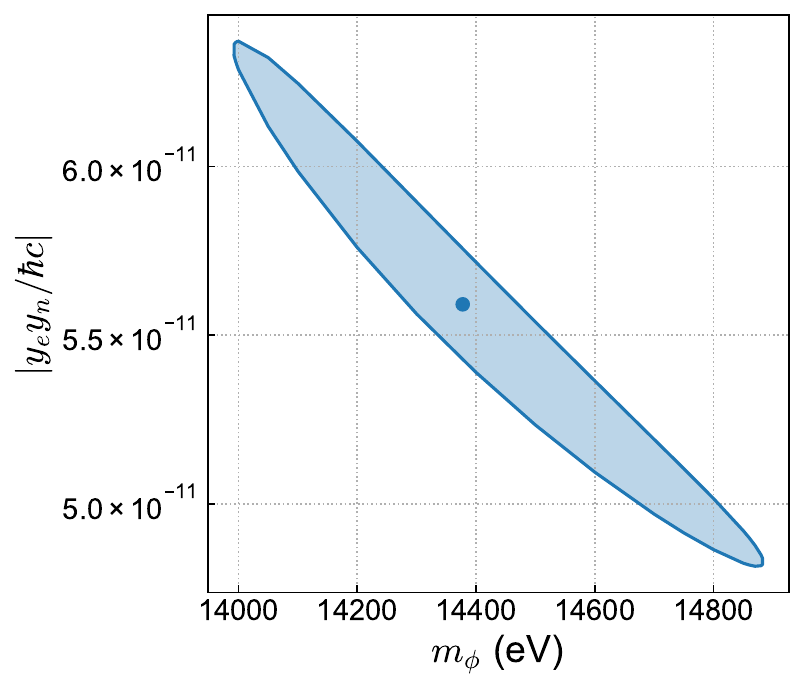}
 \caption{The parameter region of the Yukawa potential that could explain the observed 3D King nonlinearity among four transitions of 578, 431, 411, and 467. The blue point is the best value, at which $\chi^2$ reaches zero. The blue shaded area is the 95\% confidence region.} 
 \label{Fig: Combined_NP_region}
\end{figure}

Moreover, since one dof still remains, the new particle mass $m_\phi$ can be introduced as yet another fitting parameter:
\begin{eqnarray}
    && \dn{3}{AA_0} = k \w{AA_0} + f_1 \dn{1}{AA_0} + f_2 \dn{2}{AA_0} \\
    && + \alpha_{\mathrm{NP}} \left( X_3(m_\phi) - f_1 X_1(m_\phi) - f_2 X_2(m_\phi) \right) (A - A_0),  \nonumber \\
    && \dn{4}{AA_0} = k' \w{AA_0} + f_1' \dn{1}{AA_0} + f_2' \dn{2}{AA_0} \\
    && + \alpha_{\mathrm{NP}} \left( X_4(m_\phi) - f_1' X_1(m_\phi) - f_2' X_2(m_\phi) \right) (A - A_0). \nonumber
\end{eqnarray}
Then, we obtain a more refined constraint in the new physics parameter space, as depicted in Fig.~\ref{Fig: Combined_NP_region}.
The obtained favored region is incompatible with the constraint by the electron anomalous magnetic moment and the neutron scattering shown in Fig.~4c in the main text.
This discrepancy might be resolved if one considers a model of particle physics such that the contribution of the Yukawa mediator to $(g-2)_e$ is canceled by that of another new particle, e.g. a pseudoscalar particle, as proposed in \cite{Balkin2021-jr}.

It is straightforward to extend the combined fit to five or more transitions (with a given number of isotope pairs).
This enables us to test the hypothetical Yukawa potential at an arbitrary mass of the mediator since the dof becomes one or more.
Although the number of precisely measured transitions is now limited to four, we anticipate precise IS data of another transition in the near future, especially the $^1S_0 \leftrightarrow \: ^3P_2$ transition in neutral Yb atoms~\cite{Yamaguchi2010-va}.

\subsection{Individual fit of 3D King linearity with QFS-subtracted ISs}
A 3D King linearity with QFS-subtracted ISs is given by
\begin{equation}
    \sdn{3}{AA_0} = k \w{AA_0} + f_1 \sdn{1}{AA_0} + f_2 \sdn{2}{AA_0},
\end{equation}
which can be explicitly rewritten as
\begin{eqnarray}
    \dn{3}{AA_0}
    &=& k \w{AA_0} + f_1 \left( \dn{1}{AA_0} - \frac{G^{(2)}_1}{G^{(2)}_0} \dn{0}{AA_0} \right) \\
    &+& f_2 \left( \dn{2}{AA_0} - \frac{G^{(2)}_2}{G^{(2)}_0} \dn{0}{AA_0} \right) + \frac{G^{(2)}_3}{G^{(2)}_0} \dn{0}{AA_0}. \nonumber
\end{eqnarray}
Here, $\delta \nu_0$ is the IS of the reference transition $i_0$ of the subtraction, and $G^{(2)}_i / {G^{(2)}_0} \ (i=1,2,3)$ are theoretical inputs.
For a statistical test, we define $\chi^2$ as
\begin{equation}
    \chi^2 = \chi^2_{\eta} + \chi^2_{\delta \nu_0} + \chi^2_{\delta \nu_1} + \chi^2_{\delta \nu_2} + \chi^2_{\delta \nu_3}.
\end{equation}
Since we introduce 19 parameters of $\w{AA_0}$, $\dn{0}{AA_0}$, $\dn{1}{AA_0}$, $\dn{2}{AA_0}$, $k$, $f_1$, and $f_2$, the dof is given by $(5\times4) - (4\times4+3)=1$.
Using the calculated $G^{(2)}_i$ ratios (see Extended Data Table \RomanNumeralCaps{1}), we obtain $\chi^2 = 1.0 \times 10^3$ for the (431, 411, 467) transition pair with the reference transition of $i_0 = 578$.
The way of setting a constraint on the new particle is described in Methods.

\begin{table*}[!h]
    \tabcolsep = 0.2cm
    \centering
    \caption{Best-fit parameters and nonlinearities of individual 2D King plots for all transition pairs. The values in parentheses are the $1\sigma$ fitting errors. For comparison, $F_{12}$ obtained from calculated $F$ are also shown.}
    \label{tab:2D KP}
    \begin{tabular}{crrrrrr}
    \hline\hline
    Transition & \multicolumn{2}{c}{Fitting parameters} & \multicolumn{2}{c}{Nonlinearity} & \multicolumn{2}{c}{$F_{12}$ (calculation)} \\
     & $K_{12}$ (GHz) & $F_{12}$ & $\chi^2$ & Significance & Ref.~\cite{Kawasaki2024-co} & This work \\
    \hline
        (578, 431) & $37.3995172(27)$ & $-1.623073387(27)$ & $6.5\times 10^7$ & $8058 \sigma$ & $-1.834$ & $-1.856(60)$ \\ 
        (578, 411) & $-3.4527259(19)$ & $1.634717787(19)$ & $1.3\times 10^7$ & $3558 \sigma$  & $1.603$ & $1.635(43)$ \\ 
        (578, 467) & $41.0321118(54)$ & $-3.631212231(55)$ & $2.1\times 10^7$ & $4583 \sigma$  & $-3.638$ & $-3.69(12)$ \\ 
        (578, 361) & $-2.812552(85)$ & $1.33656723(92)$ & $4.6\times 10^3$ & $67 \sigma$  & $1.237$ &  \\ 
        (578, 436) & $-2.793273(95)$ & $1.6533771(10)$ & $1.7\times 10^4$ & $129 \sigma$  & $1.627$ &  \\ 
        (431, 411) & $34.2147690(19)$ & $-1.007182000(16)$ & $1.3\times 10^7$ & $3567 \sigma$  & $-0.874$ & $-0.881(28)$ \\ 
        (431, 467) & $-42.6376411(51)$ & $2.237272684(41)$ & $1.9\times 10^6$ & $1392 \sigma$  & $1.984$ & $1.990(73)$ \\ 
        (431, 361) & $27.984789(64)$ & $-0.82348707(57)$ & $8.5\times 10^3$ & $92 \sigma$  & $-0.675$ &  \\ 
        (431, 436) & $35.304484(72)$ & $-1.01867152(62)$ & $1.6\times 10^4$ & $126 \sigma$  & $-0.887$ &  \\ 
        (411, 467) & $33.3632162(49)$ & $-2.221311327(31)$ & $2.4\times 10^6$ & $1545 \sigma$  & $-2.270$ & $-2.259(71)$ \\ 
        (411, 361) & $0.010852(83)$ & $0.81761100(56)$ & $5.7\times 10^1$ & $7 \sigma$  & $0.772$ &  \\ 
        (411, 436) & $0.699789(93)$ & $1.01140685(61)$ & $1.4\times 10^2$ & $12 \sigma$  & $1.015$ &  \\ 
        (467, 361) & $12.291029(74)$ & $-0.36807578(25)$ & $2.7\times 10^3$ & $52 \sigma$  & $-0.340$ &  \\ 
        (467, 436) & $15.890863(84)$ & $-0.45531858(28)$ & $3.7\times 10^3$ & $61 \sigma$  & $-0.447$ &  \\ 
        (361, 436) & $0.68627(14)$ & $1.2370276(12)$ & $1.4\times 10^2$ & $11 \sigma$  & $1.315$ &  \\
    \hline\hline
    \end{tabular}
\end{table*}

\begin{table*}[!ht]
    \centering
    \caption{Best-fit parameters and nonlinearities of individual 3D generalized King plots for all transition pairs. The values in parentheses are the $1\sigma$ fitting errors.}
    \label{tab:3D KP}
    \begin{tabular}{crrrrr}
    \hline\hline
    Transition & \multicolumn{3}{c}{Fitting parameters} & \multicolumn{2}{c}{Nonlinearity} \\
     & $k$ (GHz) & $f_1$ & $f_2$ & $\chi^2$ & significance \\
    \hline
        (578, 431, 411) & $13.8270(48)$ & $0.88478(21)$ & $-0.46206(13)$ & $7.2 \times 10^3$ & $85 \sigma$  \\ 
        (578, 431, 467) & $-21.200(14)$ & $-0.93036(62)$ & $1.66406(38)$ & $2.6 \times 10^3$ & $51 \sigma$  \\ 
        (578, 431, 361) & $10.18(19)$ & $0.7726(84)$ & $-0.3475(52)$ & $7.2 \times 10^0$ & $2.7 \sigma$  \\ 
        (578, 431, 436) & $16.46(15)$ & $0.8177(65)$ & $-0.5149(40)$ & $2.8 \times 10^0$ & $1.7 \sigma$  \\ 
        (578, 411, 467) & $28.5771(40)$ & $2.2656(19)$ & $-3.6072(11)$ & $1.7 \times 10^3$ & $41 \sigma$  \\ 
        (578, 411, 361) & $-0.234(38)$ & $0.116(18)$ & $0.747(11)$ & $1.6 \times 10^1$ & $4.0 \sigma$  \\ 
        (578, 411, 436) & $1.052(30)$ & $-0.167(14)$ & $1.1133(86)$ & $1.1 \times 10^0$ & $1.0 \sigma$  \\ 
        (578, 467, 361) & $5.72(13)$ & $0.581(11)$ & $-0.2080(31)$ & $1.1 \times 10^1$ & $3.4 \sigma$  \\ 
        (578, 467, 436) & $9.897(98)$ & $0.5304(87)$ & $-0.3092(24)$ & $5.5 \times 10^{-2}$ & $0.2 \sigma$  \\ 
        (578, 361, 436) & $1.377(70)$ & $-0.328(33)$ & $1.483(25)$ & $9.8 \times 10^0$ & $3.1 \sigma$  \\ 
        (431, 411, 467) & $-6.617(24)$ & $1.17693(70)$ & $-1.05278(69)$ & $6.6 \times 10^2$ & $26 \sigma$  \\ 
        (431, 411, 361) & $-2.02(33)$ & $0.0598(96)$ & $0.8769(95)$ & $1.8 \times 10^1$ & $4.2 \sigma$  \\ 
        (431, 411, 436) & $3.67(25)$ & $-0.0873(74)$ & $0.9247(73)$ & $3.3 \times 10^{-1}$ & $0.6 \sigma$  \\ 
        (431, 467, 361) & $-7.69(39)$ & $1.049(20)$ & $-0.8368(91)$ & $3.1 \times 10^1$ & $5.5 \sigma$  \\ 
        (431, 467, 436) & $-2.21(30)$ & $0.950(16)$ & $-0.8798(70)$ & $5.7 \times 10^0$ & $2.4 \sigma$  \\ 
        (431, 361, 436) & $5.68(40)$ & $-0.147(12)$ & $1.059(14)$ & $9.9 \times 10^0$ & $3.1 \sigma$  \\ 
        (411, 467, 361) & $-1.69(27)$ & $0.931(18)$ & $0.0509(81)$ & $1.7 \times 10^1$ & $4.2 \sigma$  \\ 
        (411, 467, 436) & $3.17(21)$ & $0.847(14)$ & $-0.0740(63)$ & $5.3 \times 10^1$ & $0.7 \sigma$  \\ 
        (411, 361, 436) & $0.7150(27)$ & $2.15(20)$ & $-1.39(25)$ & $1.6 \times 10^1$ & $4.0 \sigma$  \\ 
        (467, 361, 436) & $4.63(29)$ & $-0.1180(86)$ & $0.917(23)$ & $7.8 \times 10^0$ & $2.8 \sigma$  \\ 
    \hline\hline
    \end{tabular}
\end{table*}

\clearpage